\documentclass[twocolumn,secnumarabic,amssymb, nobibnotes, aps, prd,nofootinbib]{revtex4-2}
\usepackage[top=1in, bottom=1in, left=0.75in, right=0.75in]{geometry}

\usepackage[...]{youngtab}
\usepackage{xfrac}
\newcommand{\mdot}{\raise1.5pt \hbox{.}}
\usepackage{appendix}
\usepackage{pgfplots}
\usepackage{amsfonts,empheq}
\usepackage{blindtext}
\usepackage{multirow}
\usepackage{adjustbox}
\usepackage{booktabs}
\usepackage{xcolor,colortbl}
\definecolor{lavender}{rgb}{0.9, 0.9, 0.98}
\usepackage{blindtext, rotating}
\usepackage[graphicx]{realboxes}
\usepackage{amssymb}
\usepackage{tabularx,multirow,array,amsmath,mathalfa}
\usepackage{physics}
\begin{document}

\title{BPS Spectra of Complex Knots}

\author{Vivek Kumar Singh}%
\thanks{Contact author}
 \email{vks2024@nyu.edu}
\affiliation{Center for Quantum and Topological Systems (CQTS), NYUAD Research Institute, New York University Abu Dhabi, PO Box 129188, Abu Dhabi, UAE  
}%

\author{Nafaa Chbili}%
\thanks{Contact author}
\email{nafaachbili@uaeu.ac.ae}
\affiliation{Department of Mathematical Sciences, United Arab Emirates University, Al Ain 15551, UAE 
}%

\begin{abstract}
Marino's conjecture remains underexplored within the framework of   $SO(N )$ string dualities. In this article, we investigated the reformulated invariants of a one-parameter family of knots $\left[{\mathcal{K}}\right]_p$ derived from tangle surgery on Manolescu's quasi-alternating knot diagrams \cite{manolescu2007}. Within topological string dualities, we have verified Marino’s integrality conjecture for these families of knots up to the Young diagram representation ${\bf R}$, with ${|\bf R|}\leq 2$. Furthermore, through our analysis, we have conjectured a closed structure for the extremal refined BPS integers for the torus knots $ \left[{\bf 3_1}\right]_{2p+1}$ and $ \left[{\bf 8_{20}}\right]_{2p+1}$, $p \in \mathbb{Z}_{\geq 0}$.  As the parameter $p$ of the knot diagram increases,  the total crossing number of a knot exceeds $16$, which we describe as a complex knot.  Interestingly, we discovered maximum number of gaps in the BPS spectra associated with complex knot families. Moreover, our observations indicated that as $p$ increases, the size of these gaps also expands.
\end{abstract}
\maketitle
\tableofcontents

\section{Introduction}
The Gopakumar-Vafa \cite{GV} and Ooguri-Vafa \cite{OV} duality conjectures connect \( U(N) \) Chern-Simons theory on \( S^3 \) to topological string theory on the resolved conifold  $\mathcal{O}(-1) + \mathcal{O}(-1)$ over $\mathbf{P}^1$. This has led to the reformulation of Chern-Simons knot polynomials, revealing integrality structures known as the Labastida-Marino-Ooguri-Vafa (LMOV) condition \cite{LM1,LMV,LM2,MV}, which count BPS states arising from M2 branes ending on M5 branes in M-theory.  There have been several attempts to address the LMOV integrality conjecture, as discussed in \cite{Liu:2007kv, Kucharski:2017ogk, kucharski2017bps}. However, the most promising result in this direction is the knot-quiver correspondence \cite{Kucharski:2017ogk}. While it does not offer a general proof of the integrality conjecture\footnote{Beyond the question of proof, the knot-quiver correspondence also represents a consistent extension of the work of Ooguri and Vafa, as highlighted in initial papers \cite{KS, kucharski2017bps}.}, it provides a significant step forward. Once a quiver that encodes the symmetric colored invariants of a specific knot is identified, one can demonstrate that all corresponding (symmetric) LMOV numbers are integers. In contrast, Sinha and Vafa explored the duality between $SO(N)$ Chern-Simons theory on $S^3$ and the closed A-model topological string theory on an orientifold-resolved conifold \cite{Sinha:2000ap}. Subsequently, Marino \cite{Marino} introduced a conjecture regarding
$SO(N)$ Wilson loop observables within the context of topological string theory, suggesting that their reformulated invariants exhibit integrality properties. This conjecture has been verified for certain torus knots and links and for  the figure-eight knot as well \cite{Marino,Stevan,PBR10,NRZ14,RS01,BR}. Recent advances  \cite{NRZ13,MMuniv,WYZDN2021} in computing colored polynomials have enabled the calculation of these integers for various knots. In \cite{MMMRVS17}, the conjecture was examined for over  $100$ prime knots, but much remains to be understood.

Our work addresses these gaps by examining Marino's conjecture for specific one-parameter knot families, denoted as $\left[\mathcal{K}\right]_p$, $p \in \mathbb{Z}$. These families are generated through tangle surgery on the quasi-alternating knot diagrams in \cite{manolescu2007}. Further,
we studied the BPS spectra for complex knots (large crossing knots) and conjectured the closed structure of extremal refined BPS integers \cite{Garoufalidis:2015ewa} of the torus knots $ \left[{\bf 3_1}\right]_{2p+1}$ and $ \left[{\bf 8_{20}}\right]_{2p+1}$, $p \in \mathbb{Z}_{\geq 0}$. Our main results confirm that the integrality conjecture is valid for these knots\footnote{Note that our results coincide for all knots with crossings less than or equal to 10, as detailed in \cite{MMMRVS17, knotebook}.}. Interestingly, we also found  maximum number of BPS gaps (\ref{BPSGAP}) in the BPS spectra for sufficiently large crossings of knots (complex knots) in the Young diagram representation of length
${|\bf R|} \leq 2$, which, to our knowledge, has not been previously observed\footnote{Recent developments have explored BPS gaps in the context of 3-manifolds \cite{Gukov:2024opc}}.

The structure of the paper is outlined as follows: In Section \ref{s.int}, we provide a review of the fundamental aspects of Chern-Simons theory to knot polynomials.
Section \ref{cbkin}, introduces the infinite family of quasi-alternating knots, for which the BPS invariants are calculated in this work.
The integrality structures in topological strings are explored in Section \ref{s5.int}. Finally,  Section \ref{conclsec}, summarizes the results obtained in this paper and suggests potential directions for future research.


\section{Chern-Simons
knot polynomials}
\label{s.int}

Chern-Simons field theory serves as a powerful tool for investigating knots and links \cite{Witten:1988hf}. This section offers a concise overview of Chern-Simons theory and its implications for knot and link invariants.

\subsection{Chern-Simons Theory}

Witten's seminal work  \cite{Witten:1988hc} laid the foundation for understanding the Jones\cite{jones1997polynomial}, respectively the  HOMFLY-PT\cite{Freyd:1985dx},  polynomial within the framework of Chern-Simons theory. These polynomials are derived by evaluating the expectation values of Wilson loop observables, which represent fundamental representations of gauge groups $SU(2)$, respectively $SU(N)$),~as extensively studied in \cite{kohno2002conformal}.
Consider pure Chern-Simons theory on the three-sphere $S^3$ for a gauge group $G$, where the action $\mathcal{S}$ is given by:
\begin{equation}
\mathcal{S}= \frac{k}{4\pi}\int_{S^3} \mathrm{Tr} \left(A \wedge dA + \frac{2}{3} A\wedge A\wedge A\right).~
\end{equation}
Here, $k$ signifies the Chern-Simons level, and `$A$' denotes the gauge connection of group $G$. The  Wilson loop operators are defined as:
\begin{align*}
W^G_{\bf R}({\mathcal{K}})&= \mathrm{Tr}_{\bf R} \left( P \exp\oint_{\mathcal{K}} A_{a}^{\mu}T^{a}_{\bf R}dx^{\mu} \right),
\end{align*}
where $P$ denotes path-ordering
of the product and $T^{a}_{\bf R}$ are the generators of the adjoint representation ${\bf R}$ of the gauge group $G$. The expectation value of these Wilson loop observables is given by:

\begin{equation}
\langle W^G_{\bf R}({\mathcal{K}}) \rangle= \frac{\int {\mathcal{D}} \!A ~~e^{\mathrm{i} \mathcal{S}} W^G_{\bf R}({\mathcal{K}})}{\int {\mathcal{D}}\! A ~~ e^{i \mathcal{S}}}~.
\end{equation}
Despite the action's independence from the metric of $S^3$, evaluating the functional integral involves an infinite-dimensional moduli space of gauge connections. However, we use tools from topological quantum field theories, following the elegant solution in \cite{Witten:1988hf}, and techniques from \cite{Kaul:1991np,Kaul:1992rs,RamaDevi:1992np} to compute knot and link invariants directly. The procedure involves decomposing $S^3$, which hosts Wilson loops carrying representations ${\bf R}_1, {\bf R}_2,\ldots, {\bf R}_p$  of the gauge group $G$, into three-manifolds with $S^2$ boundaries marked by these loops. This decomposition leads to a Chern-Simons partition function on a three-manifold with boundary $(S^2,n_1,\ldots,n_p)$, which maps to a vector in the quantum Hilbert space on the boundary. Remarkably, the basis of vector space maps into the $p$-point $\widehat{G}_k$ Wess-Zumino-Novikov-Witten(WZNW) conformal blocks. Further, the Chern-Simons quantum invariants of knots are expressed as the expectation values of Wilson loops, which involve braiding and fusion matrices on $\widehat{G}_k$ WZNW conformal blocks, providing profound insights into the topology of knots and links.

In this article, we focus on a special class of knots called quasi-alternating arborescent knots (also known as double fat graphs \cite{MMMRSS15}), utilizing braiding and fusion matrices related to four-point conformal blocks. In the following section, we shall introduce this class of links and briefly review their basic properties.
\section{Quasi non- alternating arborescent knots}\label{cbkin}
\begin{figure}[ht]
\center
{\includegraphics[width=0.47\textwidth]{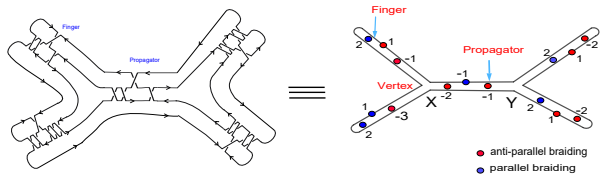}}
\caption{Arborescent knot drawn as double fat tree diagram.}\label{propa}
\end{figure}

Arborescent knots \cite{Conway70, Bonahon, caudron1982} represent a class of knots in $S^3$ which can be obtained by gluing three balls with one or more four-punctured $S^2$ boundaries as illustrated in Fig.~\ref{propa}. For clarity, almost all knots with at most  10 crossings are arborescent knots,  except the few knots listed in \cite{Mironov:2016deg}. Moreover,  these knots are represented as tree Feynman diagrams. The tree diagram representing an arborescent knot is defined as an ordinary tree with vertices of arbitrary valencies.
More precisely,  a double fat diagram consists of three parts: edges, propagators and fingers (see  Fig.\ref{propa}), more details can be found in  \cite{MMMRSS15, Mironov:2016deg}.
In the following subsections, we shall briefly outline the construction of one-parameter families of quasi-alternating non-alternating arborescent knots ($\left[\mathcal{K}\right]_p$) using tangle surgery based on the examples provided in \cite{manolescu2007}.

\subsection{ Quasi alternating knots}
\begin{figure} [h]
\begin{center}
\includegraphics[width=6cm,height=1.5cm]{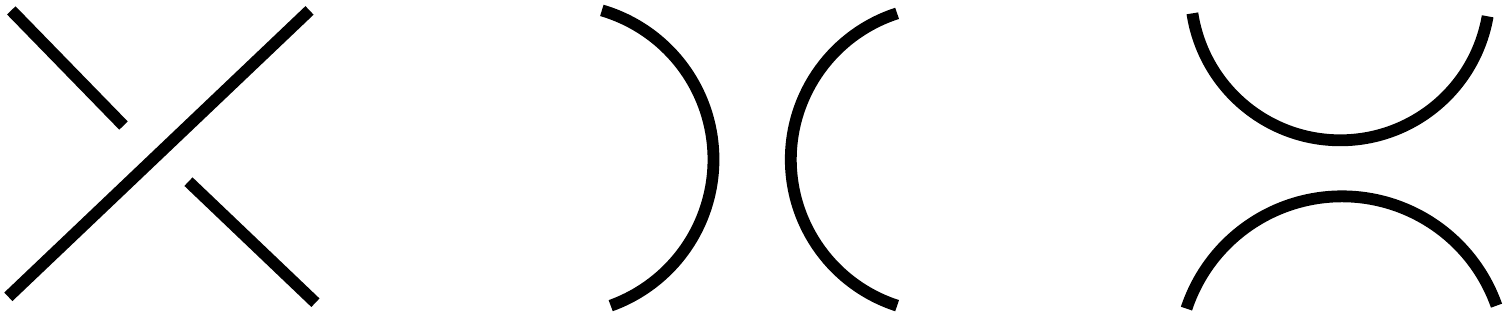} \\
{$L$}\hspace{2cm}{$L_0$}\hspace{2cm}$L_{1}$
\end{center}
\caption{The link diagram $L$ at the crossing $c$ and its smoothings $L_{0}$ and
$L_{1}$ respectively.}\label{skein}
\end{figure}

Quasi-alternating links (QA for short), introduced by Ozsváth and Szabó in the context of Heegaard Floer homology, provide a natural generalization of alternating links. The set $\mathcal{Q}$ of quasi-alternating links is defined as the smallest set of links satisfying the following properties:
\begin{itemize}
	\item The unknot belongs to $\mathcal{Q}$.
  \item If $L$ is a link with a diagram $D$ containing a crossing $c$ such that
\begin{enumerate}
\item both smoothings of the diagram $D$ at the crossing $c$, $L_{0}$ and $L_{1}$ as
in Figure \ref{skein} belong to $\mathcal{Q}$, and
\item $\det(L_{0}), \det(L_{1}) \geq 1$,
\item $\det(L) = \det(L_{0}) + \det(L_{1})$; then $L$ is in $\mathcal{Q}$ and in this
case we say $L$ is quasi-alternating at the crossing $c$ with quasi-alternating
diagram $D$.
\end{enumerate}
\end{itemize}

Where $\det(L)$ denoted the determinant of the link, and $L, L_0$ and $L_1$ represent 3 link diagrams which differ only at a small disk where they look as displayed in Figure \ref{skein}.
Notably, every non-split alternating link is QA. However, proving that a knot is QA using the definition above  is often very challenging. The first examples of QA non-alternating knots in the knot table \cite{rolfsen2003knots} are ${\bf 8_{20}}$ and ${\bf 8_{21}}$. Quasi-alternating diagrams for non-alternating knots with at most nine crossings were provided in \cite{manolescu2007}. Champanerkar and Kofman introduced a simple method for creating new QA links by replacing a QA crossing with a rational tangle of the same type in a QA diagram. We will refer to this method as ``tangle surgery" \cite{champanerkar2009twisting}\footnote{ This technique was later extended in \cite{QC15} and used for classifying quasi-alternating Montesinos links.}. In this article, we will utilize tangle surgery to generate a one-parameter family of QA non-alternating arborescent knots, which we will discuss in the following subsection.

\subsection{ QA non-alternating knots from tangle surgeries }
\begin{figure}[ht]
\center
{{\includegraphics[width=.5\textwidth]{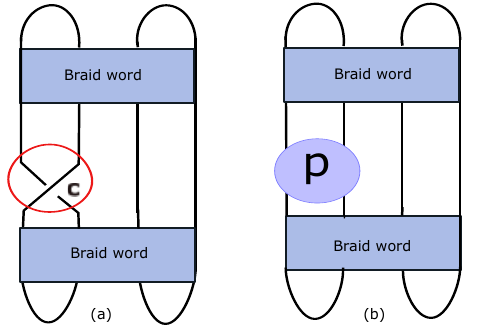}}}
\caption{An example of arborescent representation of the QA  knot \(\mathcal{K}\): (a) the QA knot structure and (b) the tangle surgery process.}\label{QAK}
\end{figure}

In this subsection, we will briefly describe how tangle surgeries are used to construct a one-parameter family of QA non-alternating arborescent knots, denoted by \(\left[\mathcal{K}\right]_p\). Let \(\mathcal{K}\) represent a QA non-alternating knot at a crossing \(c\) (indicated in red in Fig. \ref{QAK}(a)).

We now outline the steps involved in performing tangle surgeries to create this one-parameter family of quasi-alternating arborescent knots, \(\left[\mathcal{K}\right]_p\):
\begin{itemize}
    \item Consider the QA non-alternating arborescent knot diagram \(\mathcal{K}\), as given  in \cite{manolescu2007}, and identify the crossing \(c\) at which the diagram is QA (see Fig. \ref{QAK} (a)).
    \item According to \cite{champanerkar2009twisting}, the QA  nature is preserved when we perform plumbing with \(p\) copies of the same crossings (see Fig. \ref{QAK}(b)). This process generates a new family of QA non-alternating arborescent knots, denoted by
     \(\left[\mathcal{K}\right]_p\).
\end{itemize}

\begin{figure}[ht]
\center
{{\includegraphics[width=.4\textwidth]{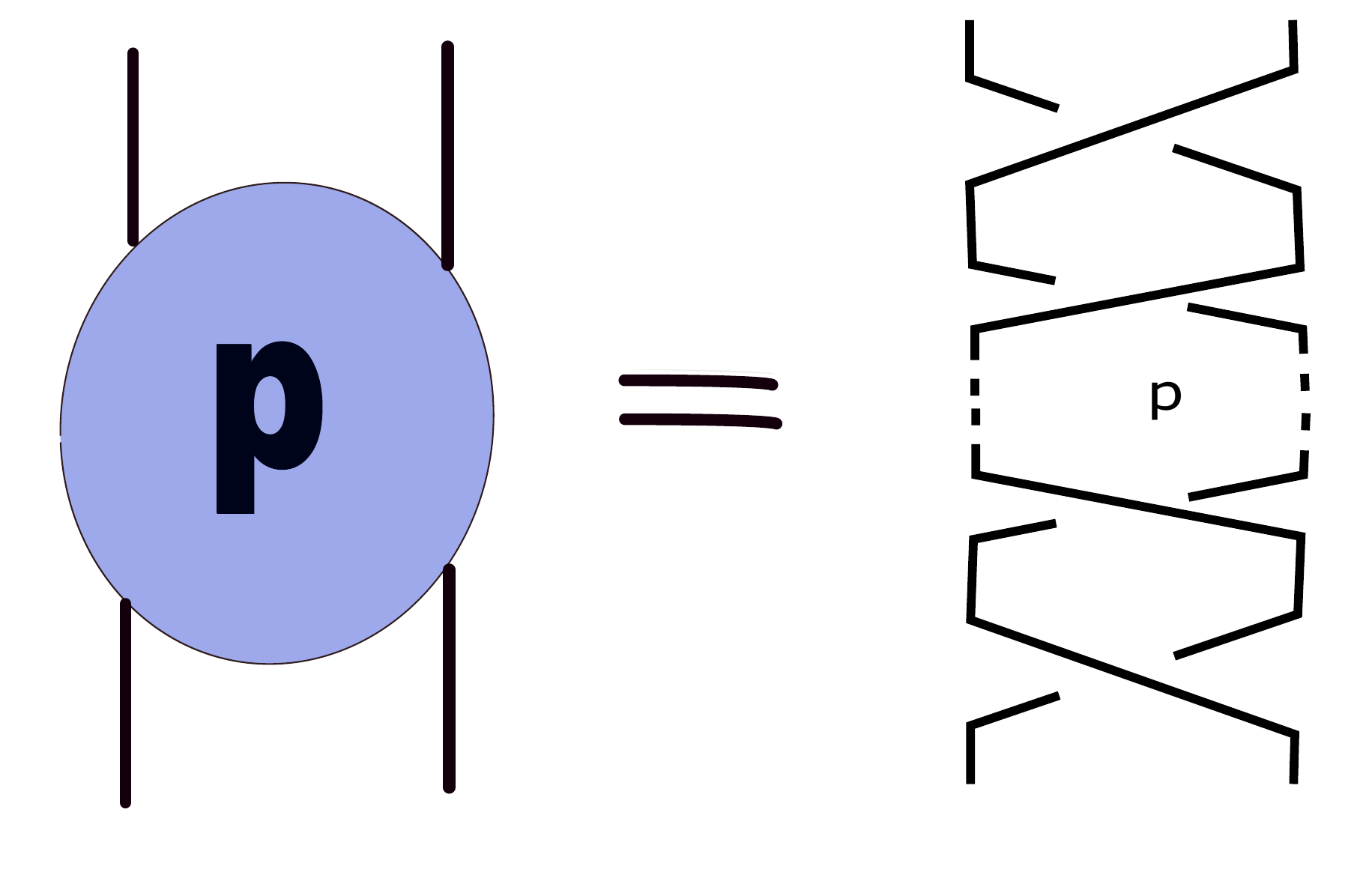}}}
\caption{Rational tangle  }\label{QAK1}
\end{figure}

To further clarify, Table \ref{table:1} presents two examples of these one-parameter families.

\begin{table}[h!]
\centering
\begin{tabular}{|c |c| c|c|}
 \hline
  knots $\left[\mathcal{K}\right]_{2p+1}$& \(\left[\mathcal{K}\right]_1\) & \(\left[\mathcal{K}\right]_3\) & \(\left[\mathcal{K}\right]_5\)  \\ [0.5ex]
 \hline
 1 & ${\bf 8_{20}}$  & ${\bf 10_{125}}$  & ${\bf 12n235}$ \\
 \hline
 2 & ${\bf 8_{21}}$ & ${\bf 10_{147}}$ & ${\bf 12n144}$ \\ [1ex]
 \hline
\end{tabular}
\caption{One-parameter families of QA non-alternating arborescent knots obtained by tangle surgeries on  ${\bf 8_{20}}$ and ${\bf 8_{21}}$.}
\label{table:1}
\end{table}

To compute the polynomial invariants for this family of arborescent knots, one must utilize quantum $6j$-symbols and braiding eigenvalues. The specific $6j$-symbols for the symmetric and anti-symmetric representations of \(SU(N)\) and \(SO(N)\) are detailed in \cite{Nawata:2013ooa, WYZDN2021}. Additionally, universal Racah matrices, which are proportional to the quantum $6j$-symbols for the adjoint representation, are provided in \cite{MMuniv}.

As a result, we have computed the colored polynomials for the class of arborescent knots discussed in this article, using these representations. The corresponding Mathematica file is attached \cite{M1-24}. For more detailed information on polynomial computations, readers can refer to \cite{Kaul:1991np, Kaul:1992rs, RamaDevi:1992np}.

These polynomials are useful for verifying the Ooguri-Vafa integrality properties and Marino's integrality conjecture within the framework of topological string duality. In the following section, we will explore their applications in the context of string dualities.
\section{Integrality structures in topological strings }\label{s5.int}
In this section, we will briefly explore the dualities of topological strings and their relationship with Chern-Simons invariants. Inspired by the AdS-CFT correspondence, Gopakumar-Vafa studied the duality between $SU(N)$ Chern-Simons theory on $S^3$ and the closed A-model topological string theory on a resolved conifold $\mathcal{O}(-1) + \mathcal{O}(-1)$ over $\mathbf{P}^1$. Specifically, they showed that the closed string partition function on the resolved conifold target space corresponds to the Chern-Simons free energy $\ln Z[S^3]$, given by:

\begin{equation}
\ln Z[S^3] = -\sum_g \mathcal{F}_{g}(t) g_{s}^{2-2g},
\end{equation}

where $\mathcal{F}_{g}(t)$ represents the genus $g$ topological string amplitude, $g_s = \frac{2 \pi}{k+N}$ denotes the string coupling constant, and $t = \frac{2 \pi i N}{k+N}$ indicates the Kähler parameter of $\mathbf{P}^1$.

Ooguri-Vafa confirmed the topological string duality conjecture with the simplest Wilson loop (unknot) observable \cite{OV} which was justified using Gopakumar-Vafa duality. Further, extended to other knots \cite{OV,LMV, LM1, LM2}, known as the LMOV conjecture, defined as:

\begin{eqnarray}
\left\langle Z(U,V)\right\rangle_{S^3} &=& \sum_{{\bf R}} \mathcal{H}^{\star}_{{\bf R}}(q,{\bf A})  \operatorname{Tr}_{{\bf R}} V, \nonumber \\
&=& \exp\left[\sum_{n=1}^{\infty} \left(\sum_{{\bf R}}\frac{1}{n} f_{{\bf R}}({\bf A}^n,q^n) \operatorname{Tr}_{{\bf R}} V^n\right)\right], \nonumber \\
&&\text{where}\nonumber \\
f_{\bf {\bf R}}(q, {\bf A}) &=& \sum_{{\bf S}, Q, s} \frac{1}{(q-q^{-1})}{\widetilde {\bf N}}_{{\bf S},Q,s}{\bf A}^Q q^{s}.\label{guv1}
\end{eqnarray}

Here, ${\bf R}$ denotes the irreducible representation of $U(N)$, ${\widetilde {\bf N}}_{{\bf S}, Q,s}$ are integers representing the number of D2-brane of bulk charge $Q$ with spin $s$ intersecting D4-brane \cite{GV1, GV2}. Using group theory methods for the powers of holonomy $V$, these reformulated invariants can be expressed in terms of colored HOMFLY-PT polynomials. For a few lower-dimensional representations(Young diagrams), their explicit forms are:

\begin{eqnarray}
f_{[1]}(q,{\bf A}) &=& {\mathcal{H}}^{*}_{[1]}(q,{\bf A}),\nonumber \\
f_{[2]}(q,{\bf A}) &=& {\mathcal{H}}^{*}_{[2]}(q,{\bf A}) - \frac{1}{2}\left({\mathcal{H}}^{*}_{[1]}(q,{\bf A})^2 + {\mathcal{H}}^{*}_{[1]}(q^2,{\bf A}^2)\right),\nonumber \\
f_{[1^2]}(q,{\bf A}) &=& {\mathcal{H}}^{*}_{[1^2]}(q,A) - \frac{1}{2}\left({\mathcal{H}}^{*}_{[1]}(q,A)^2 - {\mathcal{H}}^{*}_{[1]}(q^2,A^2)\right),\nonumber \\
\ldots \nonumber
\end{eqnarray}

where ${\cal H}^{*}_{\bf{R}}(q,{\bf A})$ is the un-normalized $\bf{R}$-colored HOMFLY-PT polynomial. Moreover, these reformulated invariants can be equivalently expressed as \cite{LMV}:
\begin{eqnarray}\label{ic}
f^{\mathcal{K}}_{{\bf R}} (q,{ A}) &=& \sum_{Q,g\ge 0,{\bf S}} C_{{\bf R}{\bf S}}\hat {\bf N}^{\mathcal{K}}_{{\bf S},Q,g}{ A}^Q (q-q^{-1})^{2g-1},
\end{eqnarray}
where $\hat {\bf N}^{\mathcal{K}}_{{\bf S},Q,g}$ are refined integers\footnote{Note that the contribution of $(q - q^{-1})^{2g-1}$ arises from the bulk of the Riemann surface with genus $g$ and the Schwinger computation\cite{OV,GV1}.} and
\begin{eqnarray*}\label{ic}
C_{{\bf R}{\bf S}} &=& \frac{1}{q-q^{-1}}\sum_{\Delta}\frac{1}{z_\Delta}\psi_{{\bf R}}(\Delta)\psi_{{\bf S}}(\Delta)\prod_{i=1}^{l(\Delta)}\left(q^{\xi_i}-q^{-\xi_i}\right).\nonumber
\end{eqnarray*}
Here,  $\psi_{{\bf R}}(\Delta)$ denotes the characters of symmetric groups, and $z_{\Delta}$ is the standard symmetric factor of the Young diagram discussed in \cite{Fulton,Mironov:2017hde}. For completeness, we have presented the reformulated invariants for a one-parameter family of QA  non-alternating knots $\left[8_{20}\right]_p$. \\
\begin{center}
\begin{tabular}{cccc}
$\hat{\bf N}^{\bf 8_{20}}_{[ 1]}:$ &
\begin{tabular}{|c|cccc|}
\hline
&&&&\\
$ g \backslash Q$ & -3 & -1 & 1 & 3 \\
&&&&\\
\hline
&&&&\\
 0&2 & -6 & 5 & -1 \\
 1&1 & -5 & 5 & -1 \\
 2&0 & -1 & 1 & 0 \\&&&&\\
\hline
\end{tabular}, $\hat{\bf N}^{\bf 10_{125}}_{[ 1]}:$ &
\begin{tabular}{|c|cccc|}
\hline
&&&&\\
$ g \backslash Q$ & -3 & -1 & 1 & 3 \\
&&&&\\
\hline
&&&&\\
0&3 & -10 & 10 & -3 \\
1& 4 & -15 & 15 & -4 \\
 2&1 & -7 & 7 & -1 \\
 3&0 & -1 & 1 & 0 \\
\hline
\end{tabular}
\end{tabular},
\end{center}
\begin{center}\begin{tabular}{cccc}
$\hat{\bf N}^{\bf 10_{12n235}}_{[ 1]}:$ &
\begin{tabular}{|c|cccc|}
\hline
&&&&\\
$ g \backslash Q$ & -3 & -1 & 1 & 3 \\
&&&&\\
\hline
&&&&\\
0&4 & -14 & 15 & -5 \\
 1&10 & -35 & 35 & -10 \\
 2&6 & -28 & 28 & -6 \\
 3&1 & -9 & 9 & -1 \\
 4&0 & -1 & 1 & 0 \\&&&&\\
\hline
\end{tabular}
\end{tabular}.
\end{center}
 Several approaches have been explored to resolve the LMOV integrality conjecture \cite{Liu:2007kv, Kucharski:2017ogk, kucharski2017bps}, with the knots-quivers correspondence \cite{Kucharski:2017ogk} emerging as the most promising. In our work, we have rigorously tested the integrality properties of the corresponding invariants ($\hat{\bf N}$) up to the level $|{\bf R}| = 2$ across various families of knots of type $\left[\mathcal{K}\right]_p$, where  $p$ is large.  More details about these results are discussed in our previous work \cite{Singh:2022jum}. Despite this, we have not observed any significant impact on the BPS structure for large $p$. \\
Additionally, Sinha-Vafa explored the duality between $SO(N)$ Chern-Simons theory on $S^3$ and closed $A$-model topological string theory on an orientifold resolved conifold. Marino conjectured that the $SO(N)$ Wilson loop observables within the framework of topological string theory, along with their reformulated invariants possess integrality properties \cite{Marino}. In this context, there exist both oriented $h_{\bf R}$ and unoriented $g_{\bf R} $ reformulated invariants, exhibiting forms resembling those conjectured by Ooguri-Vafa.

The partition function of such a topological string theory incorporates contributions from both oriented and unoriented aspects:

$$Z= {1 \over 2} Z_{\rm oriented} + Z_{\rm unoriented}~.$$

In the subsequent subsections, we will provide a brief overview of the oriented and unoriented topological string amplitudes, along with their conjectured integrality properties.
\subsection{Oriented Contribution}
Following Marino \cite{Marino}, we require $U(N)$ invariants of knots carrying composite representation $({\bf R,S})$ to incorporate orientifolding action. The composite representation ${\bf R, S}$ will have the highest weight $\Lambda_{\bf R}+\Lambda_{{\bf \bar S}}$ where $\Lambda_{\bf R}$ denotes the highest weight of representation ${\bf R}$ and $\Lambda_{{\bf \bar S}}$ denotes the highest weight of conjugate representation of ${\bf S}$. For example, $({\tiny{\yng(1),\yng(1)}})$ means adjoint representation. 
Hence the oriented contribution is given by
\begin{eqnarray}
 Z_{\rm~or}(U,V)_{\mathcal{S}^{3}}&=&\sum_t \bigg\{\sum_{{\bf R,S}} N^t_{{\bf R,S}} {\cal H}_{{\bf (R,S)}}({\bf A},q)\bigg\}Tr_{t}V \nonumber\\&=&\sum_{t}{\mathcal{R}}_{t}({\bf A},q) Tr_{t}V\nonumber\\&=&\exp\bigg[\sum\limits_{{\bf R},n}\frac{1}{n} h_{{\bf R}}({\bf A}^n,q^n) Tr_{\bf R} V^n\bigg]\nonumber,
\end{eqnarray}
where $N^t_{{\bf R,S}}$ are the Littlewood-Richardson coefficients and $h_{\bf R}$ are the oriented reformulated invariants.
The relation between $h_{\bf R}$ and colored HOMFLY-PT for this case will be:
\begin{eqnarray}
{  h}_{[1]}({\bf A},q) &=& 2{\cal H}^*_{[1]}({\bf A},q),  \nonumber \\
{  h}_{[2]}({\bf A},q) &=& 2{\cal H}^*_{[2]}({\bf A},q)+{\cal H}^*_{([1],[1])}({\bf A},q)
- 2\Big({\cal H}^*_{[1]}({\bf A},q)\Big)^2 \nonumber \\&&- {\cal H}^*_{[1]}({\bf A}^2,q^2), \nonumber\\
{  h}_{[1,1]}({\bf A},q) &=& 2{\cal H}^*_{[1,1]}({\bf A},q)+{\cal H}^*_{([1],[1])}({\bf A},q)
- 2\Big({\cal H}^*_{[1]}({\bf A},q)\Big)^2 \nonumber\\&&+{\cal H}^*_{[1]}({\bf A}^2,q^2),  \nonumber \\
\ldots  \nonumber
\end{eqnarray}
These reformulated invariants possess  integrality structures similar to $f_{\bf R}$ (\ref{ic}):

 \begin{eqnarray}
h^{\mathcal{K}}_{\bf R}(q,{\bf A})=\sum_{Q,g\ge 0,{\bf S}} C_{RQ}\hat {\bf N}^{\mathcal{K}, c=0}_{{\bf S},Q,g}{\bf A}^Q(q-q^{-1})^{2g-1},\label{hc1}
\end{eqnarray}
where $\hat {\bf N}^{c=0}_{{\bf S}, Q,g}$ are BPS integers corresponding to cross cap $c=0$\footnote{ There is a mirror symmetry under transposition of Young diagrams which relates knot polynomials as follows:
\begin{eqnarray*}
H_{{\bf R}^{tr}}(A,q) = H_{\bf R}(A,-q^{-1}).
\label{traninvers}
\end{eqnarray*}}.

\subsection{Unoriented Contribution}
The integrality structure having BPS integers corresponding to cross-caps $c=1,c=2$ from
Ooguri-Vafa operator is \cite{Marino}
\begin{eqnarray}{\label{marino}}
 Z_{\rm~unor}(U,V)_{\mathcal{S}^{3}}
&=&\sum_{{\bf R}}~~{\mathcal{F}}_{{\bf R}}({\bf A},q)Tr_{{\bf R}}V-\frac{1}{2}\sum_{{\bf R}}{\mathcal{R}}_{{\bf R}}({\bf A},q)Tr_{{\bf R}}V  \nonumber\\&=& \exp\bigg[\sum_{{\bf R} ,n}\frac{1}{n} g_{{\bf R}}({\bf A}^n,q^n) Tr_{\bf R} V^n \bigg],\nonumber
\end{eqnarray}
with the first few low-dimensional representations being
\begin{eqnarray}
{  g}_{[1]}({\bf A},q)& =& {\cal F}^*_{[1]}({\bf A},q)-{\cal H}^*_{[1]}({\bf A},q), \nonumber \\
{  g}_{[2]}({\bf A},q)& =& {\cal F}^*_{[2]}({\bf A},q)- {1\over 2}\Big({\cal F}^*_{[1]}({\bf A},q)\Big)^2-{\cal H}^*_{[2]}({\bf A},q)\nonumber \\&&
+\Big({\cal H}^*_{[1]}({\bf A},q)\Big)^2-{1\over 2}{\cal H}^*_{([1],[1])}({\bf A},q), \nonumber\\
{g}_{[1,1]}({\bf A},q) &=& {\cal F}^*_{[1,1]}({\bf A},q)- {1\over 2}\Big({\cal F}^*_{[1]}({\bf A},q)\Big)^2-{\cal H}^*_{[1,1]}({\bf A},q)\nonumber \\&&+\Big({\cal H}^*_{[1]}({\bf A},q)\Big)^2-{1\over 2}{\cal H}^*_{([1],[1])}({\bf A},q),  \nonumber\\
\ldots \nonumber
\end{eqnarray}
here ${\cal F}^*_{\bf R}({\bf A},q)$ denotes ${\bf R}$-colored Kauffman polynomials (un-normalized) and unoriented reformulated invariants $g_R$ will have contribution from both cross caps $c=1$ and $c=2$ and possess the following integrality structure:
\begin{eqnarray}
g^{\mathcal{K}}_{\bf R}(q,{\bf A})&=&\sum_{Q,g\ge 0,{\bf S}} C_{{\bf R}{\bf S}}\Big(\hat {\bf N}^{\mathcal{K},c=1}_{{\bf S},Q,g}{\bf A}^Q(q-q^{-1})^{2g}+\nonumber \\&&\hat {\bf N}^{\mathcal{K},c=2}_{{\bf S},Q,g}{\bf A}^Q(q-q^{-1})^{2g+1}\Big),\label{gc1}
\end{eqnarray}
where $\hat {\bf N}^{\mathcal{K},c=1}_{{\bf S},Q,g}$ and $\hat {\bf N}^{\mathcal{K},c=2}_{{\bf S},Q,g}$ are BPS integers corresponding to cross-caps $c=1,c=2$.
  We have verified Marino's conjectured form for the oriented and unoriented formulated invariants \cite{Marino}: $h_{[2]}, g_{[2]}, h_{[1,1]}, g_{[1,1]}$, for many one-parameter families of arborescent knots\footnote{Note that we have computed the knot polynomials for the adjoint representation, using universal duality matrices \cite{MMuniv}, these are  presented in our mathematica code \cite{M1-24}.}. While the integrality of the coefficients $\hat {\bf N}_{{\bf S},Q,g}$ has been checked for many knots in \cite{Marino,Stevan,PBR10,NRZ14,RS01,BR,MMMRVS17}, our curiosity to understand the BPS spectra as the parameter $p$ grows large led to an interesting observation: the emergence of  BPS gaps in the spectra for the representation  of length ${|\bf R|} \leq 2$. This is a non-trivial statement not observed in any literature to our knowledge. We also looked into this phenomenon across several other one-parameter knot families and found a similar pattern. We will go into more detail about our findings in the next subsection.



\subsection{BPS spectra of one-parameter knot families}
In this section, we analyze the  BPS spectra of  one-parameter families of QA non-alternating arborescent knots derived from QA non-alternating knots through tangle surgery, as illustrated in Table \ref{table:2}. Subsequently, we calculate all the necessary colored knot polynomials needed to test Marino's conjecture up to the Young diagram representation length $|2|$. The corresponding Mathematica file is attached for reference \cite{M1-24}. Using the pool of data, we have computed reformulated integers of  (\ref{hc1}) and (\ref{gc1}) for these families of knots. We have provided the Mathematica code for these computations in \cite{M1-24}. Our numerous analytical
calculations  suggest the   following: \\
\textit{ Proposition 1:}
The  refined extremal BPS invariants \footnote{ As an example, the extremal LMOV invariants are defined as \cite{Garoufalidis:2015ewa}:

\[
f^{\pm}_r(q) = \sum_j \frac{{\bf \Tilde{N}}_{r,  {\bf  Q}^\pm, j} \, q^j}{(q - q^{-1})}.
\]
Here, extremal charges \( {\bf Q}^{\pm} \) denotes the extremal powers of \(\mathbf{A}\) in (\ref{guv1}), where \( {\bf Q}^+ \) corresponds to the maximum power and \( {\bf Q}^-\) to the minimum power of \(\mathbf{A}\), and \( r \in \mathbb{N} \).}  ($\hat{\bf{N}}^{\left[\bf{3_1}\right]_{2p+1},c=2}_{\mathbf{R},{\bf Q}^{\pm},g}$) for torus knot  $\left[\bf{3_1}\right]_{2p+1}, p\in \mathbb{Z}_{\geq 0}$ (see in Table.\ref{table:3})~for young diagram representation of length $|\mathbf{R}|=2$ are as follows:\\
\begin{itemize}
\item For $\mathbf{R}=[2]$:
\begin{eqnarray*}
\hat{\bf{N}}^{\left[{\bf 3_1}\right]_{2p+1},c=2}_{[2],{\bf Q}^{-},g}&=&-  \frac{\Gamma(8+g+4p)}{\Gamma(2g+3) \Gamma(4p-g+6)}\\
&=&-\hat{\bf{N}}^{\left[{\bf 3_1}\right]_{2p+1},c=2}_{[2],{\bf Q}^{+},g}, \\
\end{eqnarray*}
where the extremal charges are $ {\bf Q}^- = 4(p+1) + 2,~ {\bf Q}^+ =2   {\bf Q}^-,~ \text{with}~  g \in \{0, 1, \ldots, 5 + 4p\} \text{ and } p \in \mathbb{Z}_{\geq 0}$.
\item For $\mathbf{R}=[1,1]$
\begin{eqnarray*}
\hat{\bf{N}}^{\left[{\bf {\bf 3_1}}\right]_{2p+1},c=2}_{[1,1],{\bf Q}^{-},g}&=& -\frac{\Gamma(7+g+4p)}{\Gamma(2g+3) \Gamma(4p-g+5)}\\ &=&-\hat{\bf{N}}^{\left[{\bf 3_1}\right]_{2p+1},c=2}_{[1,1],{\bf Q}^{+},g},
\end{eqnarray*}
\text{where} ~ ${\bf Q}^- = 4(p+1) + 2,~ {\bf Q}^+ =2  {\bf Q}^- ,~$ \text{with}~  $g \in \{0, 1, \ldots,  4p+4\}$ and $ p \in \mathbb{Z}_{\geq 0}$.
\end{itemize}
Here, \(\Gamma(n)\) is the Gamma function, defined as:
\[
\Gamma(n) = (n-1)! \quad \text{for~a~positive~integer~} n.
\]
Similarly,  the refined extremal BPS invariants for the QA non-alternating knots  $\left[{\bf 8_{20}}\right]_{2p+1},~ p \in \mathbb{Z}_{\geq 0}$ :\\
\begin{itemize}
    \item  For $\mathbf{R}=[1]$ and cross cap $c=1$:
\begin{eqnarray*}
\hat{\bf{N}}^{\left[{\bf 8_{20}}\right]_{2p+1},c=1}_{[1],{\bf Q}^{-},g} &=&  \frac{\Gamma(4+g+p)}{\Gamma(p-g+3)\Gamma(2g+2)},~\nonumber
\end{eqnarray*}
where ${\bf Q}^{-}=2 p - 6,~g \in \{0,1,2,\ldots,3+p\}.$
\item For $\mathbf{R}=[1]$ and  $c=2$:
\begin{eqnarray*}
~~\hat{\bf{N}}^{\left[{\bf 8_{20}}\right]_{2p+1},c=2}_{[1],Q,g} &=& 0 ~ \forall~ Q, ~g,~ \text{and}~ p\in \mathbb{Z}_{\geq 0}.
\end{eqnarray*}
\end{itemize}
At present, it is challenging to find a closed structure for other families and \( |\mathbf{R}| > 1 \).
Furthermore, when the parameter $p$ is sufficiently large, we refer to these knots as complex knots. 
The definition of a complex knot is as follows:\\
\textit{Complex knot:} ~A knot is referred to as a ``complex knot" when "complex" is understood as a relative term, indicating knots with more than 16 crossings, as noted in the standard tabulated atlas of 1.7 million knots \cite{Hoste1998TheF1}.

 Our numerical calculations indicate a significant BPS gap for one-parameter families of  QA  non-alternating Knot: ${\bf 8_{20},8_{21}}$, as shown in Table \ref{table:2}. The definition of BPS gap is as follows:

\textit{ Definition:}~The BPS gap of length \(n \in \mathbb{N}\) for a reformulated invariants $\hat{{\bf N}}^{\mathcal{K}}_{{\bf S},Q,g}$ of knot \(\mathcal{K}\) in the representation \(\bf S\) is a range of values of the charge \(Q\), such that
\begin{empheq}[box=\fbox]{equation}{\label{BPSGAP}}
\hat{{\bf N}}^{\mathcal{K}}_{{\bf S},Q,g} = 0 \quad \text{for} \quad Q \in [x, x+2i], \quad \forall \, g \geq 0, \, 0 \leq i\leq n ,
\end{empheq}
for some $x \in \mathbb{Z}$ and $i$ is integer.
For clarity, the complex knot $\left[\bf 8_{21}\right]_{19}$ has two BPS gaps with lengths of $2$ and $3$, as shown in the Table \ref{Table:4}. Our extensive analytical calculations indicate the following:

\textit{ Proposition 2:} The maximum number of  BPS gaps appears when the parameter  $p$ of knot families $\left[\mathcal{K}\right]_p$ exceeds a critical crossing number $d$, i.e., for  $\abs{p} \geq d\), the number of BPS gaps is maximized.
\[
\text{If } \abs{p} \geq d, \quad \#(\text{ BPS gaps}) = \text{maximal}.
\]

Further, to confirm this,  we have calculated the maximum number of BPS gaps  and a critical crossings $d$ for one-parameter  families of  QA  non-alternating knots  $\left[\mathcal{K}\right]_p$, as shown in the Tables \ref{table:2}.

\begin{table}[ht]
\centering
\begin{tabular}{|l|c|c|c|c|}
\hline
\textbf{Knot class} & \textbf{Reps} & \textbf{Cross-caps} & \textbf{$\#$ of BPS gaps} &  $d$\\
~& (${\bf R}$) & $(c)$ & ~ & ~ \\
\hline
\hline
\multirow{6}{*}{\parbox[c]{2.5cm}{ \includegraphics[width=2.5cm, height=1.5cm]{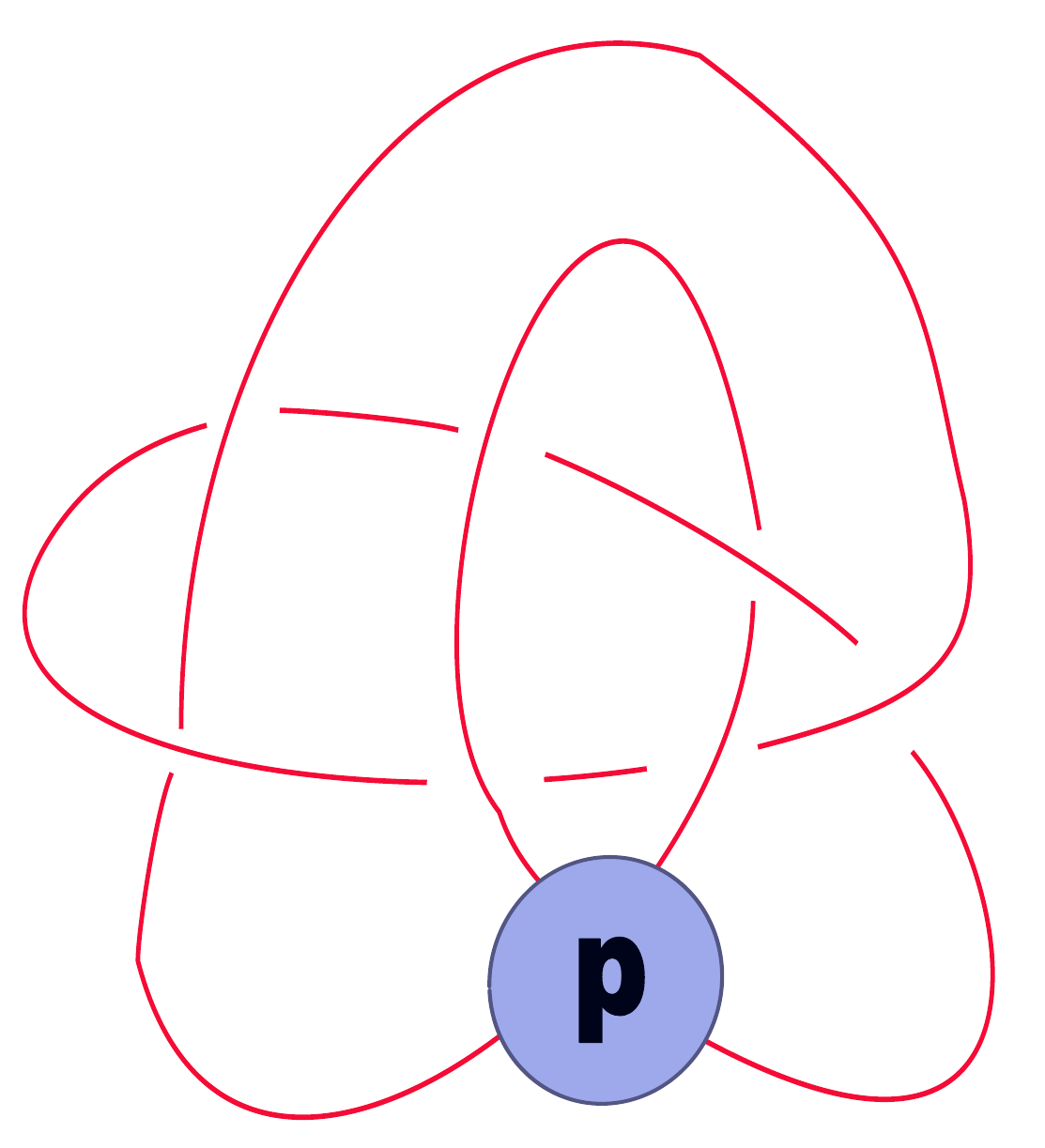}\\ $\left[{\bf 8_{21}}\right]_{2p+1}$}}
& \multirow{3}{*}{$[1]$} & 1 & 1 & 4 \\
\cline{3-5}
& & 2 & 1 & 3 \\
\cline{2-5}
& \multirow{2}{*}{$[2]$} & 1 & 2 & 7 \\
\cline{3-5}
& & 2 & 2 & 7 \\
\cline{2-5}
& \multirow{2}{*}{$[1,1]$} & 1 & 2 & 7 \\
\cline{3-5}
& & 2 & 2 & 7 \\
\hline
\multirow{6}{*}{\parbox[c]{2.5cm}{ \includegraphics[width=2.5cm, height=1.5cm]{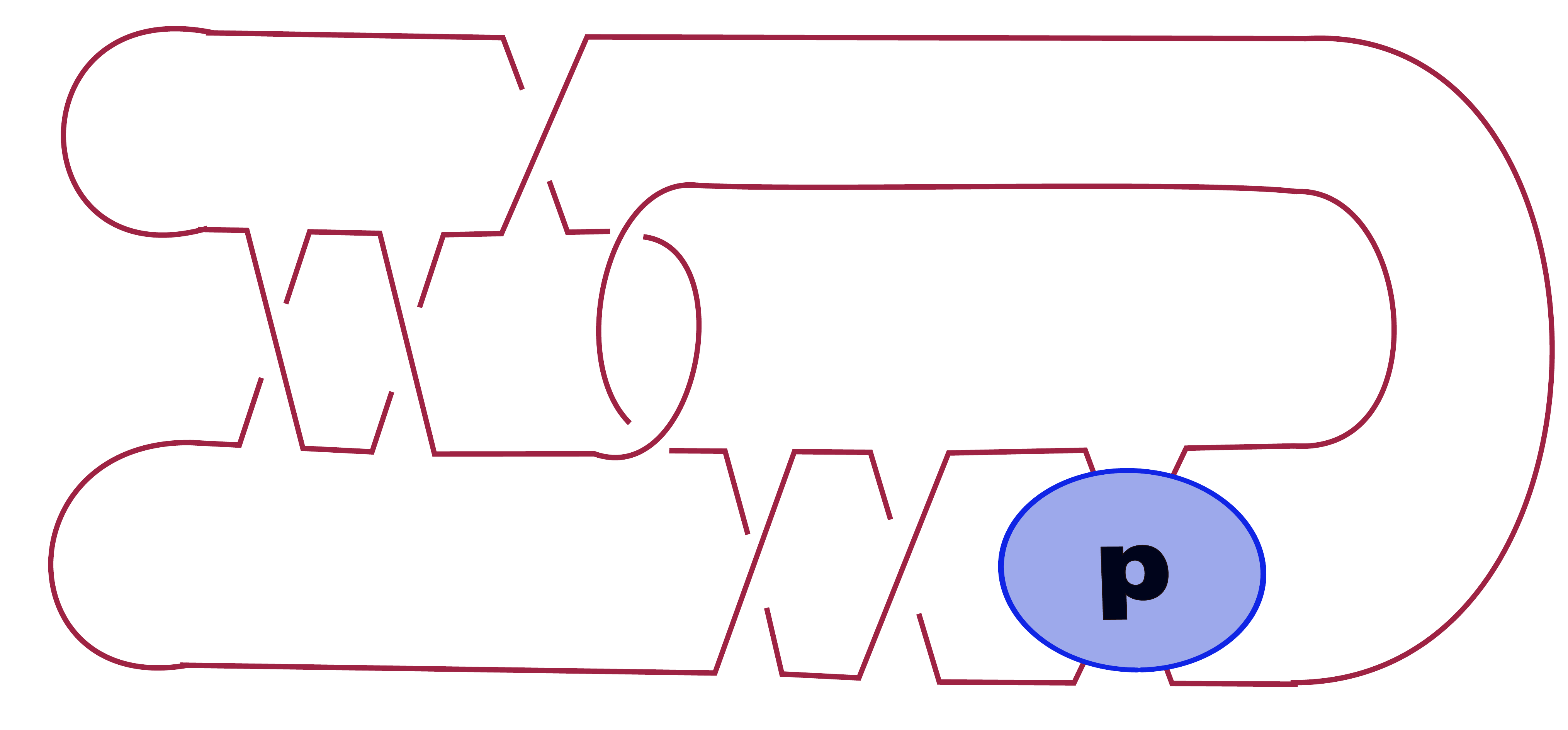}\\ $\left[{\bf 8_{20}}\right]_{2p+1}$}}
& \multirow{3}{*}{$[1]$} & 1 & 1 & 1 \\
\cline{3-5}
& & 2 & 1 & 0 \\
\cline{2-5}
& \multirow{2}{*}{$[2]$} & 1 & 1& 5 \\
\cline{3-5}
& & 2 & 2 & 4 \\
\cline{2-5}
& \multirow{2}{*}{$[1,1]$} & 1 & 1& 5 \\
\cline{3-5}
& & 2 & 2 & 4 \\
\hline
\end{tabular}
\caption{The maximum number of  gaps in the refined BPS spectra for a one-parameter family of QA non-alternating arborescent knots with \(\abs{p}\geq $d$\).}
\label{table:2}
\end{table}

For more clarity, we further investigate this phenomenon, considering random one-parameter families of arborescent knots, shown in Table \ref{table:3}, where similar gaps (not robust) were observed. It  seems that gaps may  depends on the geometry of knots.

\begin{table}[ht]
\centering
\begin{tabular}{|l|c|c|c|c|}
\hline
\textbf{Knot class} & \textbf{Reps} & \textbf{Cross-caps} & \textbf{$\#$ of BPS gaps} & $d$ \\
~& (${\bf R}$) & $(c)$ & ~ & ~ \\
\hline
\hline
\multirow{6}{*}{\parbox[c]{2.5cm}{ \includegraphics[width=2.5cm, height=1.5cm]{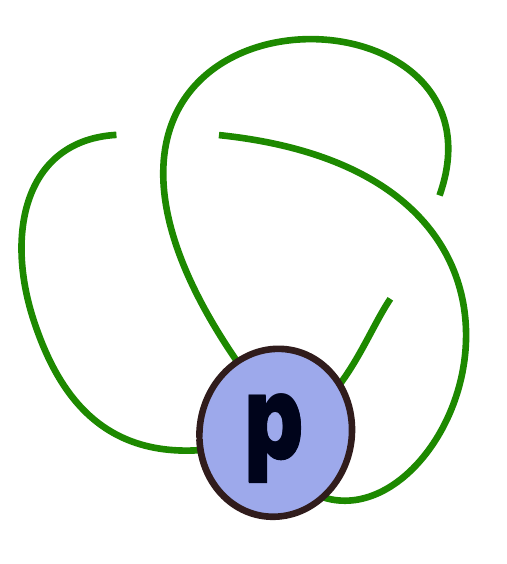}\\ $\left[{\bf 3_{1}}\right]_{2p+1}$}}
& \multirow{3}{*}{$[1]$} & 1 & 1 & 1 \\
\cline{3-5}
& & 2 & 1 & 0 \\
\cline{2-5}
& \multirow{2}{*}{$[2]$} & 1 & 1 & 3 \\
\cline{3-5}
& & 2 & 2 & 1 \\
\cline{2-5}
& \multirow{2}{*}{$[1,1]$} & 1 & 1 & 3 \\
\cline{3-5}
& & 2 & 2 & 1 \\
\hline
\multirow{6}{*}{\parbox[c]{2.5cm}{ \includegraphics[width=2.5cm, height=1.4cm]{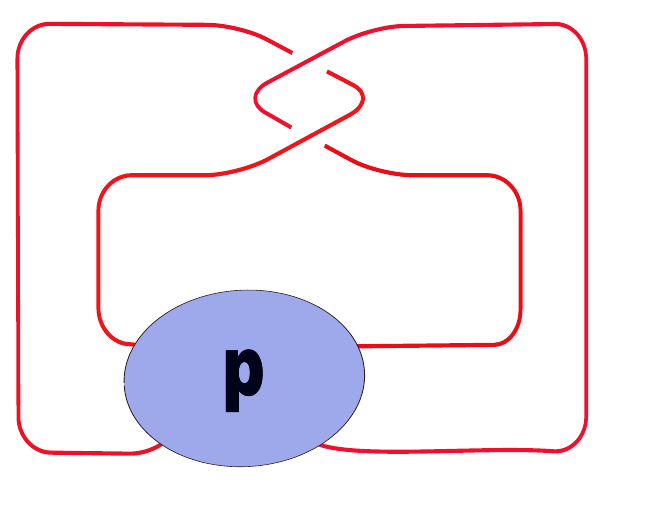}\\ Twist Knot $K_{p\in \mathbb{Z}_{>0}}$}}
& \multirow{3}{*}{$[1]$} & 1 & 1 & 3 \\
\cline{3-5}
& & 2 & 1 & 2 \\
\cline{2-5}
& \multirow{2}{*}{$[2]$} & 1 & 2 & 5 \\
\cline{3-5}
& & 2 & 2 & 6 \\
\cline{2-5}
& \multirow{2}{*}{$[1,1]$} & 1 & 2 & 5 \\
\cline{3-5}
& & 2 & 2 & 6 \\
\hline
\multirow{6}{*}{\parbox[c]{2.5cm}{ \includegraphics[width=2.5cm, height=1.4cm]{twistknot.pdf}\\ Twist Knot $K_{p\in \mathbb{Z}_{<0}}$}}
& \multirow{3}{*}{$[1]$} & 1 & 1 & 3 \\
\cline{3-5}
& & 2 & 1 & 4 \\
\cline{2-5}
& \multirow{2}{*}{$[2]$} & 1 & 2 & 6 \\
\cline{3-5}
& & 2 & 2 & 6 \\
\cline{2-5}
& \multirow{2}{*}{$[1,1]$} & 1 & 2 & 6 \\
\cline{3-5}
& & 2 & 2 & 6 \\
\hline

\end{tabular}
\caption{The maximum  number of  gaps  in the refined BPS spectra for one parameter of random arborescent knot families ($\abs{p}\geq d$).}
\label{table:3}
\end{table}
As example, we have presented the tabular structure of the reformulated invariant for complex knot $\left[{\bf 8_{21}}\right]_{13}$ in Table \ref{table:c}.

Our analysis suggests that the maximum number of BPS gaps and the critical crossing number \(d\) are primarily influenced by the cross-cap \(c\), the size of the Young diagram \(\abs{\mathbf{R}}\), and the geometry of the knots.

Based on our observations, we propose the following:\\

\textit{Conjecture 1:} Consider a one-parameter family \(\left[\mathcal{K}\right]_{2p+1}\) with \(p \in \mathbb{Z}_{\geq 0}\), where \(\mathcal{K} = \mathbf{8_{20}}, \mathbf{8_{21}}, \mathbf{3_1}\), as well as the family of twist knots \(K_p\) for  full twist \(p \in \mathbb{Z}\). Given a Young diagram representation with length \(\lvert \mathbf{R} \rvert = 2\) and cross-caps \(c = 1, 2\), there exists a  critical crossing number \(d\) such that, for \(\lvert p \rvert \geq d\), the following relation holds:
\begin{empheq}[box=\fbox]{equation}{\label{maxbps1}}
\#(\text{BPS gaps}) \text{ in } \hat{\mathbf{N}}_{[2], Q, g}^{\left[\mathcal{K}\right]_p, c} =
\#(\text{BPS gaps}) \text{ in } \hat{\mathbf{N}}_{[1,1], Q, g}^{\left[\mathcal{K}\right]_p, c}.
\end{empheq}

 Further exploration of higher-dimensional representations is necessary for a deeper understanding, though such calculations are currently challenging and will be addressed in future work.

\begin{table}[h]
\begin{tabular}{c|c|c|c|c|ccc|c|c|c}
\toprule
\text{g $\backslash$ Q} & \multicolumn{4}{c}{\textbf{Charge space}} & \multicolumn{3}{c}{\textbf{BPS gap}} & \multicolumn{3}{c}{\textbf{Charge space}} \\
\cmidrule(rl){2-5}\cmidrule(rl){6-8}\cmidrule(rl){9-11}
& 14 & 16 & 18 & 20 & 22 & 24 & 26  & 28 & 30  & 32 \\
\midrule
0 & 35 & -64 & 36 & -7 & \cellcolor{blue!30}{0} & \cellcolor{blue!30}{0} & \cellcolor{blue!30}{0} & 1 & 1 & -1     \\
1 & 314 & -610 & 359 & -63 & \cellcolor{blue!30}{0} & \cellcolor{blue!30}{0} & \cellcolor{blue!30}{0} & 1 & 1 & -2   \\
2 & 896 & -1933 & 1219 & -182 & \cellcolor{blue!30}{0} & \cellcolor{blue!30}{0} & \cellcolor{blue!30}{0} & 0 & 1 & -1  \\
3 & 1216 & -3054 & 2084 & -246 & \cellcolor{blue!30}{0} & \cellcolor{blue!30}{0} & \cellcolor{blue!30}{0} & 0 & 0 & 0  \\
4 & 917 & -2820 & 2078 & -175 & \cellcolor{blue!30}{0} & \cellcolor{blue!30}{0} & \cellcolor{blue!30}{0} & 0 & 0 & 0   \\
5 & 410 & -1634 & 1291 & -67 & \cellcolor{blue!30}{0} & \cellcolor{blue!30}{0} & \cellcolor{blue!30}{0} & 0 & 0 & 0   \\
6 & 109 & -605 & 509 & -13 & \cellcolor{blue!30}{0} & \cellcolor{blue!30}{0} & \cellcolor{blue!30}{0} & 0 & 0 & 0   \\
7 & 16 & -139 & 124 & -1 & \cellcolor{blue!30}{0} & \cellcolor{blue!30}{0} & \cellcolor{blue!30}{0} & 0 & 0 & 0   \\
8 & 1 & -18 & 17 & 0 & \cellcolor{blue!30}{0} & \cellcolor{blue!30}{0} & \cellcolor{blue!30}{0} & 0 & 0 & 0   \\
9 & 0 & -1 & 1 & 0 & \cellcolor{blue!30}{0} & \cellcolor{blue!30}{0} & \cellcolor{blue!30}{0} & 0 & 0 & 0   \\
\bottomrule
\end{tabular}
\caption{The refined BPS spectra  $\hat{\bf N}_{[1],Q,g}^{\left[{\bf 8_{21}}\right]_{13},c=1}$ of complex knot $\left[{\bf 8_{21}}\right]_{13}$.}\label{table:c}
\end{table}

\begin{table}[h]
 \begin{adjustbox}{width={8.2cm}}
\begin{tabular}{c|c|c|c|c|cc|c|c|c|c|ccc|c|c|c}
\toprule
\multicolumn{4}{c}{\textbf{Charge space}} & \multicolumn{3}{c}{\textbf{BPS gap}} & \multicolumn{3}{c}{\textbf{Charge space}}& \multicolumn{3}{c}{\textbf{BPS gap}}& \multicolumn{3}{c}{\textbf{Charge space}}  \\
\cmidrule(rl){1-4}\cmidrule(rl){5-7}\cmidrule(rl){8-11}\cmidrule(rl){12-14} \cmidrule(rl){15-17}
34 & 36 &$\ldots$ & 46 &  48& $50$ & $52$  & 54 & 56  &\ldots & 64 & 66 & 68 & 70& 72 & \ldots & 82 \\
\midrule
* & * & * & * &* & \cellcolor{blue!30}{0}&\cellcolor{blue!30}{0}  &*&* & * & *  & \cellcolor{blue!30}{0} & \cellcolor{blue!30}{0}&\cellcolor{blue!30}{0}& * & * & *   \\
*& * & * & * &*& \cellcolor{blue!30}{0} &\cellcolor{blue!30}{0}&*&* & * & *  & \cellcolor{blue!30}{0} & \cellcolor{blue!30}{0}&\cellcolor{blue!30}{0} & * & * & *  \\
*& * & * &* & *& \cellcolor{blue!30}{0}&\cellcolor{blue!30}{0} &* &* & * & * & \cellcolor{blue!30}{0} & \cellcolor{blue!30}{0}&\cellcolor{blue!30}{0} & * & * & *  \\
\vdots & \vdots & \vdots &\vdots & {\vdots}& \cellcolor{blue!30}{\vdots}&\cellcolor{blue!30}{\vdots} &\vdots & \vdots &\vdots &\vdots &\cellcolor{blue!30}{\vdots} & \cellcolor{blue!30}{\vdots}&\cellcolor{blue!30}{\vdots} & \vdots &\vdots & \vdots  \\
* & * & * &* & * &\cellcolor{blue!30}{0} & \cellcolor{blue!30}{0} &* &* & * & * &\cellcolor{blue!30}{0} & \cellcolor{blue!30}{0}&\cellcolor{blue!30}{0} & * & * & *  \\
* & *& * &* &* & \cellcolor{blue!30}{0} &\cellcolor{blue!30}{0} & *  &*& * & *  & \cellcolor{blue!30}{0} & \cellcolor{blue!30}{0}&\cellcolor{blue!30}{0} &* & * & *  \\
\bottomrule
\end{tabular}
\end{adjustbox}
\caption{The refined BPS structure  ${\hat{\bf N}}_{[2],Q,g}^{\left[{\bf 8_{21}}\right]_{17},c=2}$ of complex knot $\left[{\bf 8_{21}}\right]_{17}$. In this context, \( * \) signifies integers, while the blue region indicates BPS gaps (where all entries are zero) of lengths $2$ and $3$, respectively, whose size depends on the parameter $p$.}\label{Table:4}
\end{table}

\section{Conclusion and discussion}\label{conclsec}
In this article, we have studied one-parameter ($p$) families of QA  non-alternating arborescent knots originating from tangle surgery on Manolescu’s quasi-alternating diagrams (see Section \ref{cbkin}), with a focus on the knots \(\bf 8_{20}\) and \(\bf 8_{21}\). For large values of the parameter $p$,  we refer to these knots as `complex knots'. Further, we have computed the colored HOMFLYPT and colored Kauffman polynomials involved in equations (\ref{guv1}), (\ref{hc1}), and (\ref{gc1}) using techniques from topological quantum field theory, as outlined in \cite{Kaul:1991np,Kaul:1992rs,RamaDevi:1992np}. The detailed calculations of these polynomial invariants are provided in the accompanying Mathematica file \cite{M1-24}. Using our polynomial datum, we computed reformulated invariants for a one-parameter family of knots, denoted as \(\left[\mathcal{K}\right]_{p}\).  Despite multiple efforts to resolve the LMOV integrality conjecture \cite{Liu:2007kv, Kucharski:2017ogk, kucharski2017bps}, our main focus has been on examining the structure of the BPS spectra in the context of Marino’s Conjecture (\ref{gc1}). This conjecture has been verified for certain torus knots and links as well as other examples such as the figure-eight knot \cite{Marino,Stevan,PBR10,NRZ14,RS01,BR}. Recent advancements in computing colored polynomials \cite{NRZ13,MMuniv,WYZDN2021} have enabled the calculation of these integers for various knots. In \cite{MMMRVS17}, over 100 prime knots have been examined, though much remains to be understood. Our work addressed some of these gaps by investigating conjecture (\ref{gc1}) for the families of knots $\left[\mathcal{K}\right]_p$. We validated the integrality conjecture for these knot families up to Young diagram representations of length two, with results that coincide for all knots with $10$ or fewer crossings as reported in \cite{MMMRVS17, knotebook}. For completeness, a few examples are provided in the Appendix \ref{app}, with the full results available in the accompanying Mathematica file \cite{M1-24}.\\
Interestingly, we observed gaps (\ref{BPSGAP}) in the refined BPS spectra \(\hat{\mathbf{N}}_{{\bf R}, Q, g}^{\left[{\mathcal{K}}\right]_{p}, c}\) for configurations with \(|{\bf R}| \leq 2\). Further, we propose that the maximum number of BPS gaps occurs when the parameter \(p\) for the knot families \(\left[\mathcal{K}\right]_p\) exceeds a critical crossing number \(d\). Specifically, for \(|p| \geq d\), the number of BPS gaps reaches its maximum (refer to \textit{Proposition 2}). To quantify this, we calculated the maximum number of BPS gaps and identified the critical crossing number \(d\) for one-parameter knot families \(\left[\mathcal{K}\right]_p\), including twist knots, such as \(\mathcal{K} = \{ \mathbf{8_{20}}, \mathbf{8_{21}},{\bf 3_1}\}\), as detailed in Tables \ref{table:2} and \ref{table:3}. Our analysis indicates that for these families, a  critical crossing number \(d\) exists, when \(|p| \geq d\), the maximum number of BPS gaps satisfies the following relation \ref{maxbps1} (see \textit{conjecture 1} ). For more clarity, refer to Table \ref{Table:4} for the complex knot \(\left[{\bf 8_{21}}\right]_{17}\). Notably, as \(p \to \infty\), the lengths of these gaps also tend toward infinity; this behavior, however, has not been observed for these knots in the context of the LMOV conjecture (\ref{guv1}) within \(SU(N)\) string dualities.  Moreover, our analysis leads us to conjecture (see \textit{Proposition 1}) regarding the extremal refined BPS invariants for families of knots, specifically for torus knots such as \(\left[{\bf 3_1}\right]_{2p+1}\) and \(\left[{\bf 8_{20}}\right]_{2p+1}\), where \(p \in \mathbb{Z}_{\geq 0}\).\\
To summarize,  the appearance of  BPS gaps(\ref{BPSGAP}) seems to be influenced by the twist parameters rather than the specific types of knots selected. Additionally, it appears that the number of gaps may depend on the length of the representation $|{\bf R}|$, though we currently lack formal proof for this observation. To further clarify the BPS spectra, it will be necessary to explore higher representations ($|{\bf R}|\geq 3$), although these computations are challenging.  It would also be an interesting aspect to explore the appearance of large-size BPS gaps in the context of string dualities. The significance of this gap suggests that the $2$-homology classes labeled by $Q$ do not support BPS states. A   phenomenon that appears to be highly dependent on the knot's geometry. We hope to explore  this direction in future work.


\textbf{Acknowledgements}
We gratefully acknowledge Marcos Marino, Hisham Sati, Piotr Sulkowski,  and  P. Ramadevi for their insightful feedback on the manuscript.
 The work of NC was supported by United Arab Emirates University, UPAR grant No. G00004167. VKS’s research is funded by the ‘Tamkeen under the NYU Abu Dhabi Research Institute grant CG008 and ASPIRE Abu Dhabi under Project AARE20-336.

\bibliographystyle{apsrev4-1}

\bibliography{ref.bib}


\newpage~

\section{Reformulated Integers}{\label{app}}
In this appendix, we present several examples of reformulated invariants (\ref{gc1}) for Young diagram representations of  length $ |\mathbf{R}| = 2 $ within the one-parameter arborescent knot families $\left[\mathcal{K}\right]_{2p+1}$, where $p\in \mathbb{Z}_{\geq 0}$. This includes the knots $\left[{\bf 8_{20}}\right]_{2p+1}, \left[{\bf 8_{21}}\right]_{2p+1}, \text {the torus knot}\left[{\bf 3_1}\right]_{2p+1} $, where  $p \in \mathbb{Z}_{\geq 0}$, and the twist knot $ (K_p)_{p\in \mathbb{Z}} $ for various values of \( p \). The complete set of reformulated invariants up to $|\mathbf{R}| \leq 2$ is available in the accompanying Mathematica file \cite{M1-24}. We have verified our results for all knots with crossings less than or equal to 10, as reported in \cite{MMMRVS17, knotebook}.

\subsection{Reformulated integers for $\left[{\bf 8_{20}}\right]_{2p+1}$, $p \in \mathbb{Z}_{\geq 0}$}
For $\left[{\bf 8_{20}}\right]_{1}=\bf{8_{20}:}$\\
{

\begin{equation*}
\hat{\textbf{N}}_{[1,1]}^{{\bf 8_{20}},c=1}: \left[
\begin{array}{c|c |c|c|c|c|c|cc}
g \backslash Q&-11&-9 & -7 &-5  &-3  & -1 & 1   \\
\hline
0 &208 & -944 & 1734 & -1642 & 842 & -222 & 24 \\
 \hline
 1&2107 & -8865 & 14524 & -11641 & 4676 & -856 & 55 \\
 \hline
 2&10561 & -41834 & 61612 & -41798 & 12948 & -1529 & 40 \\
 \hline
 3&32160 & -123599 & 166318 & -95920 & 22662 & -1632 & 11 \\
 \hline
4& 64264 & -247000 & 308745 & -151799 & 26894 & -1105 & 1 \\
\hline
5& 87697 & -347013 & 408847 & -171220 & 22160 & -471 & 0 \\
\hline
6& 83551 & -350248 & 393823 & -139736 & 12731 & -121 & 0 \\
\hline
7& 56187 & -256993 & 278670 &
-82900 & 5053 & -17 & 0 \\
\hline
8& 26713 & -137658 & 145204 &- 35610 & 1352 & -1 & 0 \\
\hline
9& 8897 & -53635 & 55429 & -10923 & 232 & 0 & 0 \\
\hline
10& 2026 & -14998 & 15275 & -2326 & 23 & 0 & 0 \\
\hline
11& 300 & -2927 & 2952 & -326 & 1 & 0 & 0 \\
\hline
12& 26 & -378 & 379 & -27 & 0 & 0 & 0 \\
\hline
13& 1 & -29 & 29 & -1 & 0 & 0 & 0 \\
\hline
14 &0 & -1 & 1 & 0 & 0 & 0 & 0 \\
 \hline
\end{array}
\right];
\end{equation*}}

{\begin{equation*}
\hat{\textbf{N}}_{[1,1]}^{{\bf 8_{20}},c=2}:\left[
\begin{array}{c|c|c|c|c|c|c|c|c|c}
 g \backslash Q&-12 &-10  &-8  & -6 & -4 &-2 &0  &2 & 4 \\
 \hline
 0&315 & -1214 & 1794 &- 1260 & 438 & -100 & 36 & -10 & 1 \\
 \hline 1&
 3465 & -13225 & 18885 & -12420 & 3714 & -470 & 66 & -15 & 0 \\
 \hline 2&

 17878 & -67003 & 90314 & -53437 & 13273 & -1060 & 42 & -7 & 0 \\
 \hline 3&

 53910 & -201207 & 255638 & -133587 & 26585 & -1349 & 11 & -1 & 0 \\
 \hline 4 &

 103753 & -394200 & 474873 & -216810 & 33397 & -1014 & 1 & 0 & 0 \\
 \hline 5 &

 134083 & -531391 & 611656 & -241661 & 27769 & -456 & 0 & 0 & 0 \\
 \hline 6 &

 120036 &- 509029 & 564214 & -190755 & 15654 & -120 & 0 & 0 & 0 \\
 \hline 7 &

 75736 & -353077 & 379598 & -108244 & 6004 & -17 & 0 & 0 & 0 \\
 \hline 8 &

 33858 & -178831 & 187707 & -44274 & 1541 & -1 & 0 & 0 & 0 \\
 \hline 9 &

 10648 & -66054 & 68079 & -12926 & 253 & 0 & 0 & 0 & 0 \\
 \hline 10 &

 2301 & -17575 & 17875 & -2625 & 24 & 0 & 0 & 0 & 0 \\
 \hline 11 &

 325 &- 3277 & 3303 & -352 & 1 & 0 & 0 & 0 & 0 \\
 \hline 12 &

 27 & -406 & 407 &- 28 & 0 & 0 & 0 & 0 & 0 \\
 \hline 13 &

 1 & -30 & 30 & -1 & 0 & 0 & 0 & 0 & 0 \\
 \hline 14 &

 0 & -1 & 1 & 0 & 0 & 0 & 0 & 0 & 0 \\
 \hline

\end{array}
\right];
\end{equation*}}

\newpage ~
\newpage

{\begin{equation*}
\hat{\textbf{N}}_{[2]}^{{\bf 8_{20}},c=1}:\left[
\begin{array}{c|c|c|c|c|c|c|c|c}
g \backslash Q&
 -11 & -9 & -7 & -5 & -3 & -1 & 1 & 3 \\
 \hline
0 &
 163 & -723 & 1301 & -1217 & 631 & -179 & 25 & -1 \\ \hline
 1&
 1459 & -6030 & 9636 & -7506 & 2970 & -589 & 61 & -1 \\ \hline
 2&
 6463 & -25270 & 36235 & -23553 & 6911 & -836 & 50 & 0 \\ \hline
 3&
 17319 & -66196 & 86737 & -47418 & 10208 & -667 & 17 & 0 \\ \hline
 4&
 30172 & -116549 & 142053 & -65559 & 10204 & -323 & 2 & 0 \\ \hline
 5&
 35400 & -142768 & 164366 & -63878 & 6974 & -94 & 0 & 0 \\ \hline
 6&
 28479 & -123929 & 136524 & -44288 & 3229 & -15 & 0 & 0 \\ \hline
 7&
 15809 & -76877 & 81912 & -21832 & 989 & -1 & 0 & 0 \\ \hline
 8&
 6023 & -34068 & 35419 & -7565 & 191 & 0 & 0 & 0 \\ \hline
 9&
 1542 & -10670 & 10902 & -1795 & 21 & 0 & 0 & 0 \\ \hline
 10&
 253 & -2302 & 2325 & -277 & 1 & 0 & 0 & 0 \\ \hline
 11&
 24 & -325 & 326 & -25 & 0 & 0 & 0 & 0 \\ \hline
 12&
 1 & -27 & 27 & -1 & 0 & 0 & 0 & 0 \\ \hline
13 &
 0 & -1 & 1 & 0 & 0 & 0 & 0 & 0 \\ \hline

\end{array}
\right];
\end{equation*}}

\vspace{2cm}

{ \begin{equation*}
 \hat{\textbf{N}}_{[2]}^{{\bf 8_{20}},c=2}:\left[
\begin{array}{c|c|c|c|c|c|c|c|c|c}
g \backslash Q&
 -12 & -10 & -8 & -6 & -4 & -2 & 0 & 2 &4\\
 \hline
0 &
 248 & -954 & 1413 & -1015 & 399 & -138 & 65 & -21 & 3 \\ \hline
 1&
 2419 & -9185 & 12990 & -8445 & 2575 & -469 & 155 & -41 & 1 \\ \hline
 2&
 10970 & -40936 & 54318 & -31193 & 7435 & -708 & 143 & -29 & 0 \\ \hline
 3&
 28819 & -107691 & 134330 & -67226 & 12320 & -606 & 63 & -9 & 0 \\ \hline
 4&
 47840 & -183768 & 217177 & -93757 & 12806 & -310 & 13 & -1 & 0 \\ \hline
 5&
 52677 & -213996 & 241766 & -89041 & 8686 & -93 & 1 & 0 & 0 \\ \hline
 6&
 39561 & -175117 & 190771 & -59093 & 3893 & -15 & 0 & 0 & 0 \\ \hline
 7&
 20519 & -102241 & 108245 & -27663 & 1141 & -1 & 0 & 0 & 0 \\ \hline
 8&
 7335 & -42733 & 44274 & -9086 & 210 & 0 & 0 & 0 & 0 \\ \hline
 9&
 1772 & -12673 & 12926 & -2047 & 22 & 0 & 0 & 0 & 0 \\ \hline
 10&
 276 & -2601 & 2625 & -301 & 1 & 0 & 0 & 0 & 0 \\ \hline
 11&
 25 & -351 & 352 & -26 & 0 & 0 & 0 & 0 & 0 \\ \hline
 12&
 1 & -28 & 28 & -1 & 0 & 0 & 0 & 0 & 0 \\ \hline
 13&
 0 & -1 & 1 & 0 & 0 & 0 & 0 & 0 & 0 \\ \hline
\end{array}
\right].
\end{equation*}}
\newpage
.
\newpage
For $\left[{\bf 8_{20}}\right]_{3}=\bf{10_{125}:}$

\vspace{2cm}

\resizebox{!}{1.4cm}{$\hat{\textbf{N}}_{[2]}^{{\bf 10_{125}},c=1}:
\left[
\begin{array}{c|c|c|c|c|c|c|c|c|c|c|c|c|c|c|c|c}
Q \backslash g  & 0 & 1 & 2 & 3 & 4 & 5 & 6 & 7 & 8 & 9 & 10 & 11 & 12 & 13 & 14 &15\\
 \hline
 -7&
 251 & 2998 & 16874 & 57675 & 130570 & 204270 & 226321 & 180116 & 103590 & 42965 & 12696 & 2602 & 351 & 28 & 1 & 0 \\ \hline
-5 &
 -1237 & -13087 & -68242 & -223890 & -501659 & -798450 & -923611 & -787359 & -498241 & -234258 & -81329 & -20527 & -3656 & -435 & -31 & -1 \\ \hline
 -3&
 2601 & 23201 & 105590 & 311940 & 644634 & 963220 & 1060243 & 869287 & 533661 & 245160 & 83654 & 20853 & 3683 & 436 & 31 & 1 \\ \hline
 -1&
 -3048 & -22001 & -80999 & -197006 & -341777 & -434026 & -407512 & -283912 & -146577 & -55662 & -15298 & -2953 & -379 & -29 & -1 & 0 \\ \hline
 1&
 2090 & 12110 & 33883 & 60459 & 75794 & 69062 & 45984 & 22178 & 7605 & 1797 & 277 & 25 & 1 & 0 & 0 & 0 \\ \hline
 3&
 -641 & -2738 & -4847 & -4275 & -1394 & 813 & 1088 & 524 & 134 & 18 & 1 & 0 & 0 & 0 & 0 & 0 \\ \hline
 5&
 -203 & -1728 & -5881 & -10710 & -11768 & -8258 & -3783 & -1125 & -209 & -22 & -1 & 0 & 0 & 0 & 0 & 0 \\ \hline
 7&
 254 & 1634 & 4596 & 7111 & 6603 & 3824 & 1390 & 308 & 38 & 2 & 0 & 0 & 0 & 0 & 0 & 0 \\ \hline
 9&
 -67 & -389 & -974 & -1304 & -1003 & -455 & -120 & -17 & -1 & 0 & 0 & 0 & 0 & 0 & 0 & 0 \\ \hline
\end{array}
\right];$}

\vspace{2cm}

\resizebox{!}{1.6cm}{$\hat{\textbf{N}}_{[2]}^{{\bf 10_{125}},c=2}:\left[
\begin{array}
{c|c|c|c|c|c|c|c|c|c|c|c|c|c|c|c|c}
Q \backslash g  & 0 & 1 & 2 & 3 & 4 & 5 & 6 & 7 & 8 & 9 & 10 & 11 & 12 & 13 & 14 &15\\
 \hline
 -8&
 358 & 4746 & 28161 & 96833 & 213565 & 319241 & 334594 & 250989 & 136117 & 53382 & 14974 & 2926 & 378 & 29 & 1 & 0 \\ \hline
 -6&
 -1464 & -18542 & -106587 & -362131 & -806103 & -1243075 & -1374347 & -1112943 & -668006 & -298266 & -98603 & -23778 & -4061 & -465 & -32 & -1 \\ \hline
 -4&
 2375 & 27611 & 147405 & 469767 & 989859 & 1457070 & 1549464 & 1215184 & 710739 & 310939 & 101204 & 24129 & 4089 & 466 & 32 & 1 \\ \hline
-2 &
 -2000 & -19900 & -92400 & -257935 & -476360 & -612241 & -564352 & -379616 & -187708 & -68079 & -17875 & -3303 & -407 & -30 & -1 & 0 \\ \hline
 0&
 990 & 7685 & 27868 & 60354 & 85475 & 82763 & 56018 & 26693 & 8896 & 2026 & 300 & 26 & 1 & 0 & 0 & 0 \\ \hline
 2&
 -154 & -851 & -2445 & -4185 & -4378 & -2836 & -1136 & -273 & -36 & -2 & 0 & 0 & 0 & 0 & 0 & 0 \\ \hline
 4&
 -446 & -3008 & -8660 & -13972 & -14037 & -9220 & -4027 & -1159 & -211 & -22 & -1 & 0 & 0 & 0 & 0 & 0 \\ \hline
 6&
 570 & 3599 & 10139 & 16291 & 16349 & 10678 & 4602 & 1296 & 229 & 23 & 1 & 0 & 0 & 0 & 0 & 0 \\ \hline
 8&
 -250 & -1410 & -3565 & -5067 & -4381 & -2381 & -816 & -171 & -20 & -1 & 0 & 0 & 0 & 0 & 0 & 0 \\ \hline
 10&
 0 & 0 & 0 & 0 & 0 & 0 & 0 & 0 & 0 & 0 & 0 & 0 & 0 & 0 & 0 & 0 \\ \hline
12 &
 21 & 70 & 84 & 45 & 11 & 1 & 0 & 0 & 0 & 0 & 0 & 0 & 0 & 0 & 0 & 0 \\ \hline

\end{array}
\right];$}
\vspace{2cm}

\resizebox{!}{1.2cm}{$
\hat{\textbf{N}}_{[1,1]}^{{\bf 10_{125}},c=1}=\left[
\begin{array}{c|c|c|c|c|c|c|c|c|c|c|c|c|c|c|c|c|c}
Q \backslash g  & 0 & 1 & 2 & 3 & 4 & 5 & 6 & 7 & 8 & 9 & 10 & 11 & 12 & 13 & 14 &15&16\\
 \hline
-7 &
 324 & 4204 & 26230 & 100076 & 254552 & 451413 & 573361 & 530366 & 360583 & 180623 & 66331 & 17600 & 3278 & 406 & 30 & 1 & 0 \\ \hline
 -5&
 -1606 & -18657 & -107657 &- 392165 & -980002 &- 1751469 &- 2295410 &- 2241335 &- 1646183 &- 913122 & -381918 &- 119456 & -27461 &- 4497 &- 496 &- 33 &- 1 \\ \hline
-3 &
 3326 & 33192 & 169191 & 558143 & 1288307 & 2160042 & 2689143 & 2519990 & 1791386 & 968551 & 397193 & 122408 & 27840 & 4526 & 497 & 33 & 1 \\ \hline
 -1&
 -3711 &- 30370 &- 128542 & -358612 & -709583 &- 1026788 & -1104438 & -891104 & -541225 & -246955 & -83931 & -20878 & -3684 & -436 & -31 & -1 & 0 \\ \hline
1 &
 2335 & 15040 & 48887 & 104306 & 158043 & 174255 & 140695 & 83089 & 35631 & 10924 & 2326 & 326 & 27 & 1 & 0 & 0 & 0 \\ \hline
 3&
 -664 & -3154 & -6993 & -9613 &- 9049 & -5999 & -2776 & -869 & -174 & -20 & -1 & 0 & 0 & 0 & 0 & 0 & 0 \\ \hline
 5&
 -144 & -1033 &- 2983 & -4563 & -4118 & -2299 & -802 & -170 &- 20 &- 1 & 0 & 0 & 0 & 0 & 0 & 0 & 0 \\ \hline
7 &
 189 & 1013 & 2342 & 2925 & 2136 & 936 & 242 & 34 & 2 & 0 & 0 & 0 & 0 & 0 & 0 & 0 & 0 \\ \hline
9 &
- 49 & -235 & -475 & -497 & -286 &- 91 & -15 & -1 & 0 & 0 & 0 & 0 & 0 & 0 & 0 & 0 & 0 \\ \hline

\end{array}
\right];$}

\vspace{2cm}

\resizebox{!}{1.6cm}{
$\hat{\textbf{N}}_{[1,1]}^{{\bf 10_{125}},c=2}=\left[
\begin{array}{c|c|c|c|c|c|c|c|c|c|c|c|c|c|c|c|c|c}
Q \backslash g  & 0 & 1 & 2 & 3 & 4 & 5 & 6 & 7 & 8 & 9 & 10 & 11 & 12 & 13 & 14 &15&16\\
 \hline
-8 &
 456 & 6580 & 43456 & 168100 & 420278 & 717977 & 868394 & 760836 & 489365 & 232233 & 81029 & 20501 & 3655 & 435 & 31 & 1 & 0 \\ \hline
-6 &
 -1874 & -25929 & -165650 &- 630421 & -1580514 &- 2760968 & -3482756 & -3247157 & -2270124 & -1198484 & -477897 & -142882 & -31494 & -4961 & -528 & -34 & -1 \\ \hline
 -4&
 3042 & 39038 & 232554 & 831594 & 1974710 & 3292359 & 3991785 & 3600234 & 2448955 & 1264538 & 495472 & 146159 & 31900 & 4991 & 529 & 34 & 1 \\ \hline
 -2&
- 2504 &- 28109 &- 148181 & -470416 &- 990180 & -1457164 & -1549479 & -1215185 & -710739 & -310939 & -101204 &- 24129 & -4089 & -466 & -32 & -1 & 0 \\ \hline
 0&
 1135 & 10065 & 42757 & 109746 & 185145 & 214557 & 175253 & 102259 & 42734 & 12673 & 2601 & 351 & 28 & 1 & 0 & 0 & 0 \\ \hline
 2&
 -176 & -1180 & -3950 & -7576 & -8857 & -6578 & -3167 & -985 & -191 & -21 & -1 & 0 & 0 & 0 & 0 & 0 & 0 \\ \hline
 4&
 -331 & -1866 & -4446 & -5892 & -4799 & -2498 & -833 & -172 &- 20 & -1 & 0 & 0 & 0 & 0 & 0 & 0 & 0 \\ \hline
 6&
 418 & 2212 & 5188 & 6869 & 5582 & 2875 & 939 & 188 & 21 & 1 & 0 & 0 & 0 & 0 & 0 & 0 & 0 \\ \hline
8 &
 -181 & -846 & -1756 & -2013 & -1366 & -560 & -136 & -18 & -1 & 0 & 0 & 0 & 0 & 0 & 0 & 0 & 0 \\ \hline
10 &
 0 & 0 & 0 & 0 & 0 & 0 & 0 & 0 & 0 & 0 & 0 & 0 & 0 & 0 & 0 & 0 & 0 \\ \hline
 12&
 15 & 35 & 28 & 9 & 1 & 0 & 0 & 0 & 0 & 0 & 0 & 0 & 0 & 0 & 0 & 0 & 0 \\ \hline
\end{array}
\right],$}

 more results in \cite{M1-24}.
\newpage
.
\newpage
\subsection{Reformulated integers for $\left[{\bf 8_{21}}\right]_{2p+1}, ~ p \in \mathbb{Z}_{\geq 0}$}
For $\left[8_{21}\right]_{1}=\bf{8_{21}:}$\\
  {\small \begin{equation*}
\hat{\textbf{N}}_{[2]}^{{\bf 8_{21}},c=1}:\left[
\begin{array}{c|c|c|c|c|c|c|c|c}
g \backslash Q& 3 & 5 & 7 & 9 & 11 & 13 & 15 & 17 \\
 \hline
 0&
 -53 & 312 & -773 & 1039 & -800 & 329 & -50 & -4 \\ \hline
 1&
 -126 & 1230 & -4215 & 6782 & -4953 & 503 & 1291 & -512 \\ \hline
 2&
 -115 & 2381 & -12138 & 23006 & -10884 & -17178 & 21993 & -7065 \\ \hline
 3&
 -54 & 3007 & -23942 & 52169 & 6899 & -136406 & 142049 & -43722 \\ \hline
 4&
 -12 & 2601 & -34514 & 88766 & 91631 & -507153 & 512382 & -153701 \\ \hline
 5&
 -1 & 1519 & -36602 & 120793 & 230705 & -1149705 & 1173544 & -340253 \\ \hline
 6&
 0 & 580 & -28184 & 131898 & 324179 & -1750189 & 1824581 & -502865 \\ \hline
 7&
 0 & 137 & -15463 & 111457 & 295585 & -1881811 & 2004763 & -514668 \\ \hline
 8&
 0 & 18 & -5910 & 70107 & 184575 & -1471546 & 1595943 & -373187 \\ \hline
 9&
 0 & 1 & -1525 & 31905 & 80549 & -850516 & 933612 & -194026 \\ \hline
 10&
 0 & 0 & -252 & 10256 & 24561 & -365523 & 403423 & -72465 \\ \hline
 11&
 0 & 0 & -24 & 2258 & 5129 & -116403 & 128295 & -19255 \\ \hline
 12&
 0 & 0 & -1 & 323 & 699 & -27078 & 29606 & -3549 \\ \hline
 13&
 0 & 0 & 0 & 27 & 56 & -4468 & 4816 & -431 \\ \hline
 14&
 0 & 0 & 0 & 1 & 2 & -495 & 523 & -31 \\ \hline
 15&
 0 & 0 & 0 & 0 & 0 & -33 & 34 & -1 \\ \hline
 16&
 0 & 0 & 0 & 0 & 0 & -1 & 1 & 0 \\ \hline

\end{array}
\right];
\end{equation*}}

\begin{equation*}
\hat{\textbf{N}}_{[2]}^{{\bf 8_{21}},c=2}:\left[
\begin{array}{c|c|c|c|c|c|c|c|c|c}
g \backslash Q& 2 & 4 & 6 & 8 & 10 & 12 & 14 & 16&18 \\
 \hline

 0&
 8 & -8 & -255 & 1145 & -2190 & 2198 & -1135 & 233 & 4 \\ \hline
 1&
 6 & 29 & -1567 & 8855 & -20866 & 24110 & -12902 & 1858 & 477 \\ \hline
 2&
 1 & 56 & -4056 & 32954 & -98324 & 130245 & -70934 & 4503 & 5555 \\ \hline
 3&
 0 & 24 & -5968 & 75347 & -288055 & 436117 & -240245 & -5635 & 28415 \\ \hline
 4&
 0 & 3 & -5518 & 114663 & -569442 & 987769 & -550357 & -59574 & 82456 \\ \hline
 5&
 0 & 0 & -3288 & 119978 & -790862 & 1582048 & -891681 & -167698 & 151503 \\ \hline
 6&
 0 & 0 & -1248 & 87469 & -788878 & 1837405 & -1045648 & -275845 & 186745 \\ \hline
 7&
 0 & 0 & -289 & 44520 & -571810 & 1570361 & -898885 & -303493 & 159596 \\ \hline
 8&
 0 & 0 & -37 & 15684 & -302371 & 995221 & -570105 & -234647 & 96255 \\ \hline
 9&
 0 & 0 & -2 & 3736 & -116207 & 468289 & -266976 & -130033 & 41193 \\ \hline
 10&
 0 & 0 & 0 & 573 & -32024 & 162635 & -91767 & -51837 & 12420 \\ \hline
 11&
 0 & 0 & 0 & 51 & -6154 & 41053 & -22806 & -14721 & 2577 \\ \hline
 12&
 0 & 0 & 0 & 2 & -782 & 7312 & -3980 & -2902 & 350 \\ \hline
 13&
 0 & 0 & 0 & 0 & -59 & 870 & -462 & -377 & 28 \\ \hline
 14&
 0 & 0 & 0 & 0 & -2 & 62 & -32 & -29 & 1 \\ \hline
 15&
 0 & 0 & 0 & 0 & 0 & 2 & -1 & -1 & 0 \\ \hline

\end{array}
\right];
\end{equation*}
~
\begin{equation*}
\hat{\textbf{N}}_{[1,1]}^{{\bf 8_{21}},c=1}:\left[
\begin{array}{c|c|c|c|c|c|c|c|c}
g \backslash Q& 3 & 5 & 7 & 9 & 11 & 13 & 15 & 17 \\
 \hline
 0&
 -37 & 220 &- 553 & 757 &- 594 & 247 & -36 & -4 \\ \hline
 1&
 -75 & 728 &- 2590 & 4299 & -3081 & -2 & 1149 & -428 \\ \hline
 2&
 -52 & 1157 &- 6493 & 12799 &- 4677 & -13828 & 16241 & -5147 \\ \hline
 3&
 -13 & 1195 &- 11346 & 25939 & 10282 & -89465 & 91348 &- 27940 \\ \hline
 4&
 -1 & 837 & -14465 & 40281 & 56661 &- 286031 & 288632 & -85914 \\ \hline
 5&
 0 & 382 & -13303 & 50365 & 112245 & -562272 & 577670 & -165087 \\ \hline
 6&
 0 & 106 & -8616 & 49537 & 129640 & -741179 & 780099 & -209587 \\ \hline
 7&
 0 & 16 & -3826 & 36396 & 97300 & -686014 & 738053 & -181925 \\ \hline
 8&
 0 & 1 & -1127 & 19163 & 49374 & -457436 & 500149 & -110124 \\ \hline
 9&
 0 & 0 &- 209 & 7019 & 17098 & -222418 & 245346 & -46836 \\ \hline
 10&
 0 & 0 & -22 & 1733 & 3983 & -78911 & 87137 & -13920 \\ \hline
 11&
 0 & 0 & -1 & 274 & 597 & -20197 & 22155 & -2828 \\ \hline
 12&
 0 & 0 & 0 & 25 & 52 &- 3629 & 3926 & -374 \\ \hline
 13&
 0 & 0 & 0 & 1 & 2 &- 434 & 460 &- 29 \\ \hline
 14&
 0 & 0 & 0 & 0 & 0 & -31 & 32 &- 1 \\ \hline
 15&
 0 & 0 & 0 & 0 & 0 & -1 & 1 & 0 \\ \hline

\end{array}
\right];
\end{equation*}
\newpage
.
\newpage

\begin{equation*}
\hat{\textbf{N}}_{[1,1]}^{{\bf 8_{21}},c=2}:\left[
\begin{array}{c|c|c|c|c|c|c|c|c|c}
g \backslash Q& 2 & 4 & 6 & 8 & 10 & 12 & 14 & 16&18 \\
 \hline
 0&
 10 &- 16 & -182 & 864 & -1690 & 1732 & -910 & 188 & 4 \\ \hline
 1&
 7 & 1 & -942 & 5762 &- 14208 & 16875 &- 9096 & 1206 & 395 \\ \hline
 2&
 1 & 14 &- 2019 & 18622 & -59300 & 81222 &- 44388 & 1922 & 3926 \\ \hline
 3&
 0 & 3 & -2433 & 36965 &- 153783 & 242518 &- 134027 &- 6576 &17333 \\ \hline
 4&
 0 & 0 & -1804 & 48360 &- 267559 & 488002 &- 273019 &- 37442 & 43462 \\ \hline
 5&
 0 & 0 &- 827 & 42737 & -323727 & 689016 &- 390109 & -85664 & 68574 \\ \hline
 6&
 0 & 0 & -225 & 25688 & -277462 & 697810 & -398609 & -118989 & 71787 \\ \hline
 7&
 0 & 0 & -33 & 10447 & -169835 & 513078 & -294178 & -110803 & 51324 \\ \hline
 8&
 0 & 0 & -2 & 2817 & -74195 & 275130 & -157361 & -71771 & 25382 \\ \hline
 9&
 0 & 0 & 0 & 481 &- 22884 & 107249 &- 60799 & -32713 & 8666 \\ \hline
 10&
 0 & 0 & 0 & 47 & -4857 & 29995 & -16750 & -10438 & 2003 \\ \hline
 11&
 0 & 0 & 0 & 2 & -674 & 5854 &- 3202 & -2279 & 299 \\ \hline
 12&
 0 & 0 & 0 & 0 &- 55 & 756 & -403 & -324 & 26 \\ \hline
13 &
 0 & 0 & 0 & 0 & -2 & 58 & -30 & -27 & 1 \\ \hline
 14&
 0 & 0 & 0 & 0 & 0 & 2 &- 1 & -1 & 0 \\ \hline
\end{array}
\right];
\end{equation*}
For $\left[{\bf 8_{21}}\right]_{3}=\bf{10_{147}:}$\\
$\hat{\textbf{N}}_{[2]}^{{\bf 10_{147}},c=1}:$
\begin{equation*}
\left[
\begin{array}{c|c|c|c|c|c|c|c|c|c|c}
g \backslash Q &7&9&11&13&15&17&19&21&23&25\\
 \hline
0 &
 -401 & 2217 & -4933 & 5458 & -2788 & 118 & 462 & -131 & 0 & -2 \\ \hline
 1&
 -2103 & 15142 & -40705 & 50650 & -25970 & -354 & 2269 & 1720 & -182 & -467 \\ \hline
 2&
 -4707 & 48875 & -168489 & 241415 & -120641 & 4219 & -58468 & 74523 & -5845 & -10882 \\ \hline
 3&
 -5957 & 98908 & -460396 & 763468 & -331240 & 119167 & -864721 & 858069 & -65855 & -111443 \\ \hline
 4&
 -4734 & 140464 & -923895 & 1764769 & -450690 & 792158 & -5583238 & 5309923 & -394809 & -649948 \\ \hline
 5&
 -2449 & 147385 & -1428773 & 3149583 & 311435 & 2759279 & -22030654 & 21012441 & -1490745 & -2427502 \\ \hline
 6&
 -821 & 116188 & -1731172 & 4501680 & 2978298 & 5938571 & -59481804 & 57746106 & -3863445 & -6203601 \\ \hline
 7&
 -171 & 68576 & -1643027 & 5261555 & 7394602 & 8250936 & -116436053 & 115627837 & -7212063 & -11312192 \\ \hline
 8&
 -20 & 29890 & -1213241 & 5053578 & 11387102 & 7038988 & -171129590 & 173954449 & -9977760 & -15143396 \\ \hline
 9&
 -1 & 9418 & -690957 & 3958642 & 12378114 & 2412120 & -193213035 & 200733916 & -10408755 & -15179462 \\ \hline
 10&
 0 & 2075 & -300285 & 2493519 & 9944001 & -2334943 & -170179033 & 180193295 & -8272535 & -11546094 \\ \hline
 11&
 0 & 302 & -98149 & 1242542 & 6023720 & -4302267 & -118112888 & 126999599 & -5035617 & -6717242 \\ \hline
 12&
 0 & 26 & -23611 & 481483 & 2773220 & -3562119 & -64965783 & 70643573 & -2349621 & -2997168 \\ \hline
 13&
 0 & 1 & -4040 & 142305 & 969920 & -1939833 & -28366450 & 31057861 & -836988 & -1022776 \\ \hline
 14&
 0 & 0 & -464 & 31293 & 255594 & -752996 & -9808530 & 10764846 & -225322 & -264421 \\ \hline
 15&
 0 & 0 & -32 & 4938 & 49858 & -213019 & -2666696 & 2920782 & -44987 & -50844 \\ \hline
 16&
 0 & 0 & -1 & 527 & 6974 & -43769 & -562593 & 612351 & -6449 & -7040 \\ \hline
 17&
 0 & 0 & 0 & 34 & 661 & -6376 & -90128 & 97099 & -627 & -663 \\ \hline
 18&
 0 & 0 & 0 & 1 & 38 & -625 & -10586 & 11247 & -37 & -38 \\ \hline
 19&
 0 & 0 & 0 & 0 & 1 & -37 & -859 & 897 & -1 & -1 \\ \hline
 20&
 0 & 0 & 0 & 0 & 0 & -1 & -43 & 44 & 0 & 0 \\ \hline
 21&
 0 & 0 & 0 & 0 & 0 & 0 & -1 & 1 & 0 & 0 \\ \hline
\end{array}
\right];
\end{equation*}

\newpage
.
\newpage

$\hat{\textbf{N}}_{[2]}^{{\bf 10_{147}},c=2}:$ {\small
\begin{equation*}
\left[
\begin{array}{c|c|c|c|c|c|c|c|c|c|c|c}
g \backslash Q &6&8&10&12&14&16&18&20&22&24&26\\

 \hline
 0&
 88 & -286 & -648 & 3578 & -4486 & 154 & 3859 & -2867 & 539 & 65 & 4 \\
  \hline
 1&
 286 & -890 & -8583 & 41416 & -53973 & -11202 & 79887 & -59630 & 11421 & 383 & 885 \\ \hline
 2&
 358 & -846 & -41630 & 227970 & -334454 & -150192 & 749714 & -561843 & 97748 & -4646 & 17821 \\ \hline
 3&
 229 & 155 & -111993 & 787617 & -1376003 & -896887 & 4217712 & -3144325 & 429240 & -58101 & 152356 \\ \hline
 4&
 79 & 875 & -194816 & 1892021 & -4073143 & -3308514 & 16034032 & -11766619 & 947013 & -260901 & 729973 \\ \hline
 5&
 14 & 691 & -234846 & 3305863 & -8967613 & -8454244 & 43907799 & -31444595 & 267025 & -611532 & 2231438 \\ \hline
 6&
 1 & 251 & -202459 & 4293570 & -14957693 & -15769449 & 89911921 & -62503804 & -4692428 & -757154 & 4677244 \\ \hline
 7&
 0 & 44 & -126342 & 4193647 & -19140393 & -22047200 & 140913400 & -94912001 & -15647339 & -258250 & 7024434 \\ \hline
 8&
 0 & 3 & -57024 & 3099447 & -18952724 & -23432675 & 171575910 & -112061100 & -28660621 & 713528 & 7775256 \\ \hline
 9&
 0 & 0 & -18371 & 1736113 & -14593108 & -19077041 & 163902254 & -104044419 & -35772721 & 1407464 & 6459829 \\ \hline
 10&
 0 & 0 & -4104 & 734446 & -8747550 & -11935355 & 123591101 & -76465893 & -32622117 & 1376932 & 4072540 \\ \hline
 11&
 0 & 0 & -602 & 232370 & -4071158 & -5734994 & 73784737 & -44606932 & -22442638 & 881847 & 1957370 \\ \hline
 12&
 0 & 0 & -52 & 53976 & -1460713 & -2105835 & 34873937 & -20641209 & -11833154 & 396919 & 716131 \\ \hline
 13&
 0 & 0 & -2 & 8914 & -398933 & -584526 & 13004205 & -7544701 & -4811051 & 128345 & 197749 \\ \hline
 14&
 0 & 0 & 0 & 989 & -81219 & -120311 & 3796433 & -2160144 & -1505991 & 29757 & 40486 \\ \hline
 15&
 0 & 0 & 0 & 66 & -11914 & -17771 & 856211 & -477760 & -359622 & 4837 & 5953 \\ \hline
 16&
 0 & 0 & 0 & 2 & -1188 & -1779 & 145989 & -79846 & -64296 & 524 & 594 \\ \hline
 17&
 0 & 0 & 0 & 0 & -72 & -108 & 18174 & -9735 & -8329 & 34 & 36 \\ \hline
 18&
 0 & 0 & 0 & 0 & -2 & -3 & 1557 & -816 & -738 & 1 & 1 \\ \hline
 19&
 0 & 0 & 0 & 0 & 0 & 0 & 82 & -42 & -40 & 0 & 0 \\ \hline
 20&
 0 & 0 & 0 & 0 & 0 & 0 & 2 & -1 & -1 & 0 & 0 \\ \hline
\end{array}
\right];
\end{equation*}}

{\small \begin{equation*}
\hat{\textbf{N}}_{[1,1]}^{{\bf 10_{147}},c=1}:\left[
\begin{array}{c|c|c|c|c|c|c|c|c|c|c}
g \backslash Q &7&9&11&13&15&17&19&21&23&25\\

 \hline
0 &
 -285 & 1615 & -3669 & 4122 & -2106 & 42 & 392 & -109 & 0 &- 2 \\ \hline
 1&
 -1264 & 9591 & -26814 & 34222 & -17576 & -432 & 1065 & 1785 & -168 & -409 \\ \hline
 2&
 -2324 & 26776 & -98825 & 146833 & -72607 & 4705 & -51577 & 60256 & -4702 & -8535 \\ \hline
 3&
 -2351 & 47080 & -243186 & 421284 & -170597 & 90475 & -631880 & 615013 & -47046 & -78792 \\ \hline
 4&
 -1453 & 58504 & -443401 & 889107 & -153933 & 508305 & -3619556 & 3430128 & -252886 & -414815 \\ \hline
 5&
 -568 & 53682 & -623470 & 1456471 & 390143 & 1533369 & -12840242 & 12285368 & -859306 & -1395447 \\ \hline
 6&
 -136 & 36520 & -681378 & 1916994 & 1792674 & 2844040 & -31261201 & 30552823 & -2002830 & -3197506 \\ \hline
 7&
 -18 & 18161 &- 575206 & 2059681 & 3648771 & 3301328 & -55127655 & 55220090 & -3348697 & -5196455 \\ \hline
 8&
 -1 & 6456 & -371308 & 1802990 & 4832446 & 2097660 & -72763419 & 74671477 & -4121229 & -6155072 \\ \hline
 9&
 0 & 1587 & -181175 & 1268395 & 4552049 & -25971 & -73436771 & 77024699 & -3789923 & -5412890 \\ \hline
 10&
 0 & 255 & -65843 & 703987 & 3156123 & -1509192 & -57473056 & 61389663 & -2625928 & -3576009 \\ \hline
 11&
 0 & 24 & -17445 & 302281 & 1633866 & -1681910 & -35179093 & 38102192 & -1374974 & -1784941 \\ \hline
 12&
 0 & 1 & -3258 & 98327 & 633561 & -1085905 & -16906475 & 18479474 & -542872 & -672853 \\ \hline
 13&
 0 & 0 & -405 & 23618 & 182958 & -478334 & -6374552 & 6997067 & -160237 & -190115 \\ \hline
 14&
 0 & 0 & -30 & 4040 & 38725 & -150033 & -1874777 & 2056412 & -34748 & -39589 \\ \hline
 15&
 0 & 0 & -1 & 464 & 5828 & -33666 & -424915 & 463546 & -5366 & -5890 \\ \hline
 16&
 0 & 0 & 0 & 32 & 590 & -5297 & -72691 & 78516 & -558 & -592 \\ \hline
 17&
 0 & 0 & 0 & 1 &36 &- 556 & -9069 & 9659 & -35 & -36 \\ \hline
 18&
 0 & 0 & 0 & 0 & 1 & -35 & -778 & 814 & -1 & -1 \\ \hline
 19&
 0 & 0 & 0 & 0 & 0 & -1 & -41 & 42 & 0 & 0 \\ \hline
 20&
 0 & 0 & 0 & 0 & 0 & 0 & -1 & 1 & 0 & 0 \\ \hline
\end{array}
\right];
\end{equation*}
}
\newpage
~
\newpage
$\hat{\textbf{N}}_{[1,1]}^{{\bf 10_{147}},c=2}:$ { \small \begin{equation*}
\left[
\begin{array}{c|c|c|c|c|c|c|c|c|c|c|c}
g \backslash Q &6&8&10&12&14&16&18&20&22&24&26\\
 \hline
 0&
 66 &-202 & -552 & 2853 &-3523 & -45 &3351 & -2480 & 474 & 54 & 4 \\ \hline
 1&
 175 & -504 & -6020 & 29099 & -38234 & -9944 & 61722 & -46090 & 8825 & 200 & 771 \\ \hline
 2&
 159 & -304 & -24860 & 142745 & -216192 & -108578 & 523312 & -392256 & 66506 & -4228 & 13696 \\ \hline
 3&
 66 & 219 & -57680 & 443308 & -816167 & -571835 & 2681732 & -1995049 & 251959 & -40561 & 104008 \\ \hline
 4&
 13 & 363 & -86814 & 958132 & -2213130 & -1895575 & 9315310 &- 6799393 & 429585 & -152637 & 444146 \\ \hline
 5&
 1 & 178 & -90064 & 1498070 & -4438880 & -4370656 & 23297362 &- 16540217 & -270878 & -294631 & 1209715 \\ \hline
 6&
 0 & 38 & -65972 & 1724919 & -6697180 & -7335907 & 43439099 & -29858741 & -3192127 & -266171 & 2252042 \\ \hline
 7&
 0 & 3 & -34294 & 1476240 &- 7689083 & -9165867 & 61685542 & -41017865 & -8293003 & 50713 & 2987614 \\ \hline
 8&
 0 & 0 &-12529 & 942701 & -6766832 & -8623148 & 67621366 & -43575095 & -12928987 & 442800 & 2899724 \\ \hline
 9&
 0 & 0 & -3132 & 448474 & -4578703 & -6141805 & 57713671 & -36147618 & -13977031 & 593351 & 2092793 \\ \hline
 10&
 0 & 0 & -508 & 157650 & -2378931 & -3315196 & 38533368 & -23529094 & -11060248 & 460278 & 1132681 \\ \hline
 11&
 0 & 0 & -48 & 40253 &- 943396 & -1351293 & 20153725 & -12027265 & -6571567 & 239245 & 460346 \\ \hline
 12&
 0 & 0 & -2 & 7238 & -282193 & -411993 & 8238650 &- 4813115 & -2965167 & 86976 & 139606 \\ \hline
 13&
 0 & 0 & 0 & 867 & -62407 & -92278 & 2614948 & -1496721 & -1017766 & 22269 & 31088 \\ \hline
 14&
 0 & 0 & 0 & 62 & -9870 & -14711 & 636485 & -356951 &- 263891 & 3945 & 4931 \\ \hline
 15&
 0 & 0 & 0 & 2 & -1054 & -1578 & 116362 & -63910 & -50810 & 461 & 527 \\ \hline
 16&
 0 & 0 & 0 & 0 &- 68 & -102 & 15442 & -8299 & -7039 & 32 & 34 \\ \hline
 17&
 0 & 0 & 0 & 0 & -2 & -3 & 1403 & -737 & -663 & 1 & 1 \\ \hline
 18&
 0 & 0 & 0 & 0 & 0 & 0 & 78 & -40 & -38 & 0 & 0 \\ \hline
 19&
 0 & 0 & 0 & 0 & 0 & 0 & 2 & -1 & -1 & 0 & 0 \\ \hline
\end{array}
\right].
\end{equation*}
}
\subsection{Reformulated integers for Twist Knot $\left[K\right]_{p}$,~ with full twist $p \in \mathbb{Z}$}

$\bullet$ For $p < 0,~\left[K\right]_{-1}: \bf{3_1}$

$\hat{\textbf{N}}_{[2]}^{{\bf 3_{1}},c=1}:\left[
\begin{array}{c|c|c|c|c|c}
g \backslash Q &3&5&7&9&11\\
 \hline
 0&
 -16 & 69 & -111 & 79 & -21 \\ \hline
 1&
 -20 & 146 & -307 & 251 & -70 \\ \hline
 2&
 -8 & 128 & -366 & 330 & -84 \\ \hline
 3&
 -1 & 56 & -230 & 220 & -45 \\ \hline
 4&
 0 & 12 & -79 & 78 & -11 \\ \hline
 5&
 0 & 1 & -14 & 14 & -1 \\ \hline
 6&
 0 & 0 & -1 & 1 & 0 \\ \hline
\end{array}
\right];$

$\hat{\textbf{N}}_{[2]}^{{\bf 3_{1}},c=2}:\left[
\begin{array}{c|c|c|c|c}
g \backslash Q &6&8&10&12\\

 \hline
0 &
 -21 & 63 & -63 & 21 \\ \hline
 1&
 -70 & 231 & -231 & 70 \\ \hline
 2&
 -84 & 322 & -322 & 84 \\ \hline
 3&
 -45 & 219 & -219 & 45 \\ \hline
 4&
 -11 & 78 & -78 & 11 \\ \hline
 5&
 -1 & 14 & -14 & 1 \\ \hline
 6&
 0 & 1 & -1 & 0 \\ \hline
\end{array}
\right];$

$\hat{\textbf{N}}_{[1,1]}^{{\bf 3_{1}},c=1}:\left[
\begin{array}{c|c|c|c|c|c}
g \backslash Q &3&5&7&9&11\\

 \hline
 0&
 -8 & 39 &- 69 & 53 &- 15 \\ \hline
 1&
 -6 & 61 & -146 & 126 &- 35 \\ \hline
 2&
 -1 & 37 & -128 & 120 &- 28 \\ \hline
 3&
 0 & 10 &- 56 & 55 &- 9 \\ \hline
 4&
 0 & 1 &- 12 & 12 & -1 \\ \hline
 5&
 0 & 0 & -1 & 1 & 0 \\ \hline
\end{array}
\right];$

$\hat{\textbf{N}}_{[1,1]}^{{\bf 3_{1}},c=2}:\left[
\begin{array}{c|c|c|c|c}
g \backslash Q &6&8&10&12\\

 \hline
0 &
 -15 & 45 &- 45 & 15 \\ \hline
 1&
- 35 & 120 & -120 & 35 \\ \hline
 2&
 -28 & 119 & -119 & 28 \\ \hline
 3&
 -9 & 55 & -55 & 9 \\ \hline
 4&
 -1 & 12 & -12 & 1 \\ \hline
 5&
 0 & 1 & -1 & 0 \\ \hline
\end{array}
\right].$

\newpage~
\newpage~

$\bullet \left[K\right]_{-2}$: $\bf{5_2}$\\

$\hat{\textbf{N}}_{[2]}^{{\bf 5_{2}},c=1}:$
\begin{equation*}
\left[
\begin{array}{c|c|c|c|c|c|c|c}
g \backslash Q &3&5&7&9&11&13&15\\

 \hline
0 &
 -4 & -5 & -9 & 189 & -403 & 324 & -92 \\ \hline
1 &
 -5 & -5 & -250 & 1765 & -3599 & 2935 & -841 \\ \hline
 2&
 -1 & -1 & -1022 & 6993 & -14931 & 12490 & -3528 \\ \hline
 3&
 0 & 0 & -1948 & 15525 & -36254 & 31272 & -8595 \\ \hline
 4&
 0 & 0 & -2111 & 21471 & -56421 & 50238 & -13177 \\ \hline
 5&
 0 & 0 & -1389 & 19476 & -59047 & 54157 & -13197 \\ \hline
 6&
 0 & 0 & -562 & 11868 & -42651 & 40138 & -8793 \\ \hline
 7&
 0 & 0 & -136 & 4879 & -21490 & 20656 & -3909 \\ \hline
 8&
 0 & 0 & -18 & 1332 & -7525 & 7353 & -1142 \\ \hline
 9&
 0 & 0 & -1 & 231 & -1793 & 1773 & -210 \\ \hline
 10&
 0 & 0 & 0 & 23 & -277 & 276 & -22 \\ \hline
 11&
 0 & 0 & 0 & 1 & -25 & 25 & -1 \\ \hline
 12&
 0 & 0 & 0 & 0 & -1 & 1 & 0 \\ \hline
\end{array}
\right];
\end{equation*}

$\hat{\textbf{N}}_{[2]}^{{\bf 5_{2}},c=2}:$
\begin{equation*}
    \left[
\begin{array}{c|c|c|c|c|c|c|c}
g \backslash Q &4&6&8&10&12&14&16\\

 \hline
0 &
 -10 & 48 & -45 & -229 & 627 & -567 & 176 \\ \hline
 1&
 -18 & 125 & 125 & -2600 & 6154 & -5396 & 1610 \\ \hline
 2&
 -8 & 120 & 1106 & -10891 & 25676 & -22478 & 6475 \\ \hline
 3&
 -1 & 55 & 2444 & -23924 & 59960 & -53231 & 14697 \\ \hline
 4&
 0 & 12 & 2704 & -31680 & 87935 & -79717 & 20746 \\ \hline
 5&
 0 & 1 & 1728 & -27105 & 86086 & -79821 & 19111 \\ \hline
 6&
 0 & 0 & 665 & -15502 & 58140 & -55066 & 11763 \\ \hline
 7&
 0 & 0 & 152 & -5985 & 27474 & -26504 & 4863 \\ \hline
 8&
 0 & 0 & 19 & -1540 & 9065 & -8875 & 1331 \\ \hline
 9&
 0 & 0 & 1 & -253 & 2046 & -2025 & 231 \\ \hline
 10&
 0 & 0 & 0 & -24 & 301 & -300 & 23 \\ \hline
 11&
 0 & 0 & 0 & -1 & 26 & -26 & 1 \\ \hline
 12&
 0 & 0 & 0 & 0 & 1 & -1 & 0 \\ \hline
\end{array}
\right];
\end{equation*}

$\hat{\textbf{N}}_{[1,1]}^{{\bf 5_{2}},c=1}:$
\begin{equation*}
\left[
\begin{array}{c|c|c|c|c|c|c|c}
g \backslash Q &3&5&7&9&11&13&15\\

\hline
0 &
 -2 &- 3 &- 13 & 151 &- 315 & 256 & -74 \\ \hline
 1&
 -1 &- 2 &- 179 & 1186 &- 2420 & 1985 &- 569 \\ \hline
 2&
 0 & 0 &- 568 & 3975 &- 8653 & 7296 & -2050 \\ \hline
 3&
 0 & 0 & -863 & 7442 & -18041 & 15736 &- 4274 \\ \hline
 4&
 0 & 0 & -735 & 8582 &- 23882 & 21548 & -5513 \\ \hline
 5&
 0 & 0 & -366 & 6370 &- 20957 & 19490 &- 4537 \\ \hline
 6&
 0 & 0 &- 105 & 3090 & -12445 & 11869 & -2409 \\ \hline
 7&
 0 & 0 & -16 & 971 &- 5016 & 4879 &- 818 \\ \hline
 8&
 0 & 0 & -1 & 190 & -1350 & 1332 & -171 \\ \hline
 9&
 0 & 0 & 0 & 21 &- 232 & 231 & -20 \\ \hline
 10&
 0 & 0 & 0 & 1 & -23 & 23 & -1 \\ \hline
 11&
 0 & 0 & 0 & 0 &- 1 & 1 & 0 \\ \hline

\end{array}
\right];
\end{equation*}

$\hat{\textbf{N}}_{[1,1]}^{{\bf 5_{2}},c=2}:$
\begin{equation*}
\left[
\begin{array}{c|c|c|c|c|c|c|c}
g \backslash Q &4&6&8&10&12&14&16\\

 \hline
 0&
 -5 & 27 &- 20 &- 193 & 504 &- 454 & 141 \\ \hline
 1&
 -6 & 54 & 135 & -1794 & 4185 &- 3659 & 1085 \\ \hline
 2&
 -1 & 36 & 671 & -6229 & 14806 &- 12986 & 3703 \\ \hline
 3&
 0 & 10 & 1111 & -11340 & 29296 &- 26181 & 7104 \\ \hline
 4&
 0 & 1 & 934 &- 12353 & 36190 &- 33133 & 8361 \\ \hline
 5&
 0 & 0 & 442 &- 8566 & 29512 &- 27680 & 6292 \\ \hline
 6&
 0 & 0 & 119 &- 3876 & 16320 &- 15639 & 3076 \\ \hline
 7&
 0 & 0 & 17 & -1140 & 6156 & -6003 & 970 \\ \hline
 8&
 0 & 0 & 1 & -210 & 1560 &- 1541 & 190 \\ \hline
 9&
 0 & 0 & 0 &- 22 & 254 &- 253 & 21 \\ \hline
 10&
 0 & 0 & 0 & -1 & 24 &- 24 & 1 \\ \hline
 11&
 0 & 0 & 0 & 0 & 1 & -1 & 0 \\ \hline
\end{array}
\right].
\end{equation*}

$\bullet$ For $p > 0, ~
\left[K\right]_{1}: \bf{4_1}$\\

$\hat{\textbf{N}}_{[2]}^{{\bf 4_{1}},c=1}:\left[
\begin{array}{c|c|c|c|c|c|c}
g \backslash Q &-5&-3&-1&1&3&5\\

 \hline
0 &
 5 & -19 & 34 & -38 & 25 & -7 \\ \hline
 1&
 10 & -40 & 75 & -99 & 75 & -21 \\ \hline
 2&
 6 & -29 & 57 & -98 & 85 & -21 \\ \hline
 3&
 1 & -9 & 18 & -47 & 45 & -8 \\ \hline
 4&
 0 & -1 & 2 & -11 & 11 & -1 \\ \hline
 5&
 0 & 0 & 0 & -1 & 1 & 0 \\ \hline
\end{array}
\right];$

\vspace{1cm}
$\hat{\textbf{N}}_{[2]}^{{\bf 4_{1}},c=2}:\left[
\begin{array}{c|c|c|c|c|c|c|c}
g \backslash Q &-6&-4&-2&0&2&4&6\\

 \hline
0 &
 10 & -32 & 40 & -45 & 65 & -55 & 17 \\ \hline
 1&
 15 & -56 & 65 & -75 & 145 & -130 & 36 \\ \hline
 2&
 7 & -36 & 38 & -44 & 128 & -121 & 28 \\ \hline
 3&
 1 & -10 & 10 & -11 & 56 & -55 & 9 \\ \hline
 4&
 0 & -1 & 1 & -1 & 12 & -12 & 1 \\ \hline
 5&
 0 & 0 & 0 & 0 & 1 & -1 & 0 \\ \hline
\end{array}
\right];$

\vspace{1cm}
$\hat{\textbf{N}}_{[1,1]}^{{\bf 4_{1}},c=1}:\left[
\begin{array}{c|c|c|c|c|c|c}
g \backslash Q &-5&-3&-1&1&3&5\\

 \hline
 0&
 7 & -25 & 38 &- 34 & 19 &- 5 \\ \hline
 1&
 21 &- 75 & 99 & -75 & 40 &- 10 \\ \hline
 2&
 21 &- 85 & 98 & -57 & 29 & -6 \\ \hline
 3&
 8 &- 45 & 47 &- 18 & 9 & -1 \\ \hline
 4&
 1 &- 11 & 11 &- 2 & 1 & 0 \\ \hline
 5&
 0 &- 1 & 1 & 0 & 0 & 0 \\ \hline
\end{array}
\right];$

\vspace{1cm}
$\hat{\textbf{N}}_{[1,1]}^{{\bf 4_{1}},c=2}:\left[
\begin{array}{c|c|c|c|c|c|c|c}
g \backslash Q &-6&-4&-2&0&2&4&6\\
 \hline
0 &
 17 & -55 & 65 & -45 & 40 &- 32 & 10 \\ \hline
 1&
 36 & -130 & 145 &- 75 & 65 & -56 & 15 \\ \hline
 2&
 28 & -121 & 128 &- 44 & 38 & -36 & 7 \\ \hline
 3&
 9 & -55 & 56 & -11 & 10 &- 10 & 1 \\ \hline
 4&
 1 & -12 & 12 &- 1 & 1 &- 1 & 0 \\ \hline
 5&
 0 &- 1 & 1 & 0 & 0 & 0 & 0 \\ \hline
\end{array}
\right];$\\
\newpage
$\left[K\right]_{2}: \bf{6_1}$\\

$\hat{\textbf{N}}_{[2]}^{{\bf 6_{1}},c=1}:$
\begin{equation*}
\left[
\begin{array}{c|c|c|c|c|c|c|c|c}
g \backslash Q &-9&-7&-5&-3&-1&1&3&5\\
\hline
0& 40 & -143 & 191 & -101 & -17 & 52 & -26 & 4 \\\hline
1&
 335 & -1180 & 1484 & -754 & 14 & 169 & -78 & 10 \\\hline
2&
 1113 & -3974 & 4724 & -2146 & 132 & 231 & -86 & 6 \\\hline
3&
 1994 & -7442 & 8443 & -3335 & 216 & 168 & -45 & 1 \\\hline
4&
 2122 & -8582 & 9374 & -3135 & 166 & 66 & -11 & 0 \\ \hline
5&
 1390 & -6370 & 6748 & -1846 & 66 & 13 & -1 & 0 \\ \hline
6&
 562 & -3090 & 3196 & -682 & 13 & 1 & 0 & 0 \\
 \hline
7&
 136 & -971 & 987 & -153 & 1 & 0 & 0 & 0 \\
 \hline
8&
 18 & -190 & 191 & -19 & 0 & 0 & 0 & 0 \\
 \hline
9&
 1 & -21 & 21 & -1 & 0 & 0 & 0 & 0 \\
 \hline
10&
 0 & -1 & 1 & 0 & 0 & 0 & 0 & 0 \\\hline
\end{array}
\right];
\end{equation*}

$\hat{\textbf{N}}_{[2]}^{{\bf 6_{1}},c=2}$:
{\small \begin{equation*}
\left[
\begin{array}{c|c|c|c|c|c|c|c|c|c}
g \backslash Q &-10&-8&-6&-4&-2&0&2&4&6\\
 \hline
0&
 110 & -358 & 409 & -174 & -42 & 123 & -74 & -11 & 17 \\\hline
1&
 705 & -2405 & 2783 & -1240 & 40 & 331 & -231 & -19 & 36 \\\hline
2&
 2017 & -7167 & 8163 & -3415 & 312 & 379 & -309 & -8 & 28 \\\hline
3&
 3214 & -12119 & 13449 & -5003 & 431 & 232 & -212 & -1 & 9 \\\hline
4&
 3080 & -12730 & 13742 & -4368 & 273 & 79 & -77 & 0 & 1 \\\hline
5&
 1834 & -8672 & 9128 & -2380 & 90 & 14 & -14 & 0 & 0 \\\hline
6&
 681 & -3892 & 4012 & -816 & 15 & 1 & -1 & 0 & 0 \\\hline
7&
 153 & -1141 & 1158 & -171 & 1 & 0 & 0 & 0 & 0 \\\hline
8&
 19 & -210 & 211 & -20 & 0 & 0 & 0 & 0 & 0 \\
 \hline
9&
 1 & -22 & 22 & -1 & 0 & 0 & 0 & 0 & 0 \\\hline
10&
 0 & -1 & 1 & 0 & 0 & 0 & 0 & 0 & 0 \\\hline
\end{array}
\right];
\end{equation*}}

$\hat{\textbf{N}}_{[1,1]}^{{\bf 6_{1}},c=1}:$
\begin{equation*}
\left[
\begin{array}{c|c|c|c|c|c|c|c|c}
g \backslash Q &-9&-7&-5&-3&-1&1&3&5\\
\hline 0&
 50 & -179 & 239 & -133 & 3 & 34 & -16 & 2 \\ \hline
1& 495 & -1753 & 2221 & -1167 & 149 & 87 & -36 & 4 \\\hline 2&
 1965 & -6987 & 8403 &- 3968 & 526 & 88 &- 28 & 1 \\\hline 3&
 4229 & -15524 & 17866 & -7441 & 834 & 45 & -9 & 0 \\\hline 4&
 5502 & -21471 & 23815 &- 8582 & 726 & 11 &- 1 & 0 \\\hline 5&
 4536 & -19476 & 20944 &- 6370 & 365 & 1 & 0 & 0 \\\hline 6&
 2409 & -11868 & 12444 & -3090 & 105 & 0 & 0 & 0 \\\hline 7&
 818 & -4879 & 5016 &- 971 & 16 & 0 & 0 & 0 \\\hline 8&
 171 & -1332 & 1350 & -190 & 1 & 0 & 0 & 0 \\\hline 9&
 20 & -231 & 232 &- 21 & 0 & 0 & 0 & 0 \\\hline 10&
 1 & -23 & 23 & -1 & 0 & 0 & 0 & 0 \\\hline 11&
 0 & -1 & 1 & 0 & 0 & 0 & 0 & 0 \\\hline
\end{array}
\right];\end{equation*}

$\hat{\textbf{N}}_{[1,1]}^{{\bf 6_{1}},c=2}:$
{\small \begin{equation*}
\left[
\begin{array}{c|c|c|c|c|c|c|c|c|c}
g \backslash Q &-10&-8&-6&-4&-2&0&2&4&6\\
\hline 0&
147 &- 479 & 542 &- 221 & -28 & 78 &- 43 &- 6 & 10 \\\hline 1&
 1096 &- 3725 & 4309 & -1894 & 140 & 176 & -111 & -6 & 15 \\\hline 2&
 3709 & -13041 & 14939 & -6340 & 685 & 155 &- 113 & -1 & 7 \\\hline 3&
 7105 &- 26199 & 29358 & -11394 & 1118 & 65 &- 54 & 0 & 1 \\\hline 4&
 8361 &- 33135 & 36203 &- 12365 & 935 & 13 & -12 & 0 & 0 \\\hline 5&
 6292 &- 27680 & 29513 & -8567 & 442 & 1 & -1 & 0 & 0 \\\hline 6&
 3076 &- 15639 & 16320 & -3876 & 119 & 0 & 0 & 0 & 0 \\\hline 7&
 970 &- 6003 & 6156 &- 1140 & 17 & 0 & 0 & 0 & 0 \\\hline 8&
 190 & -1541 & 1560 &- 210 & 1 & 0 & 0 & 0 & 0 \\\hline 9&
 21 &- 253 & 254 &- 22 & 0 & 0 & 0 & 0 & 0 \\\hline 10&
 1 & -24 & 24 & -1 & 0 & 0 & 0 & 0 & 0 \\\hline 11&
 0 & -1 & 1 & 0 & 0 & 0 & 0 & 0 & 0 \\\hline
\end{array}
\right].~~
\end{equation*}}
For more results, see \cite{M1-24}.
\subsection{Reformulated integers for Torus Knot $\left[{\bf 3_1}\right]_{2p+1}, p \in \mathbb{Z}_{\geq 0}$}
For $\left[{\bf 3_1}\right]_{3}: \bf{5_1}$\\

$\hat{\textbf{N}}_{[2]}^{{\bf 5_{1}},c=1}:$
{\small \begin{equation*}
\left[
\begin{array}{c|c|c|c|c|c|c}
g \backslash Q &7&9&11&13&15&17\\
 \hline
 0&
 -120 & 415 & -415 & -45 & 275 & -110 \\ \hline
 1&
 -490 & 2085 & -2085 & -1215 & 2750 & -1045 \\ \hline
 2&
 -819 & 4663 & -4663 & -6364 & 11110 & -3927 \\ \hline
 3&
 -724 & 5994 & -5994 & -15644 & 24090 & -7722 \\ \hline
 4&
 -365 & 4822 & -4822 & -22372 & 31746 & -9009 \\ \hline
 5&
 -105 & 2500 & -2500 & -20370 & 27118 & -6643 \\ \hline
 6&
 -16 & 833 & -833 & -12307 & 15503 & -3180 \\ \hline
 7&
 -1 & 172 & -172 & -4998 & 5985 & -986 \\ \hline
 9&
 0 & 20 & -20 & -1349 & 1540 & -191 \\ \hline
 10&
 0 & 1 & -1 & -232 & 253 & -21 \\ \hline
 11&
 0 & 0 & 0 & -23 & 24 & -1 \\ \hline
 12&
 0 & 0 & 0 & -1 & 1 & 0 \\ \hline
\end{array}
\right];
\end{equation*}}

$\hat{\textbf{N}}_{[2]}^{{\bf 5_{1}},c=2}:$
\begin{equation*}
\left[
\begin{array}{c|c|c|c|c|c|c}
g \backslash Q &10&12&14&16&18&20\\

 \hline
 0&
 -55 & 0 & 275 & -275 & 0 & 55 \\ \hline
 1&
 -495 & 0 & 2750 & -2750 & 0 & 495 \\ \hline
 2&
 -1716 & 0 & 11110 & -11110 & 0 & 1716 \\ \hline
 3&
 -3003 & 0 & 24090 & -24090 & 0 & 3003 \\ \hline
 4&
 -3003 & 0 & 31746 & -31746 & 0 & 3003 \\ \hline
 5&
 -1820 & 0 & 27118 & -27118 & 0 & 1820 \\ \hline
 6&
 -680 & 0 & 15503 & -15503 & 0 & 680 \\ \hline
 7&
 -153 & 0 & 5985 & -5985 & 0 & 153 \\ \hline
 9&
 -19 & 0 & 1540 & -1540 & 0 & 19 \\ \hline
 10&
 -1 & 0 & 253 & -253 & 0 & 1 \\ \hline
 11&
 0 & 0 & 24 & -24 & 0 & 0 \\ \hline
 12&
 0 & 0 & 1 & -1 & 0 & 0 \\ \hline
\end{array}
\right];
\end{equation*}

{\small $\hat{\textbf{N}}_{[1,1]}^{{\bf 5_{1}},c=1}:$
 \begin{equation*}
\left[
\begin{array}{c|c|c|c|c|c|c}
g \backslash Q &7&9&11&13&15&17\\
 \hline
0 &
 -80 & 285 & -285 & -55 & 225 & -90 \\
  \hline
 1&
 -260 & 1190 & -1190 & -910 & 1875 & -705 \\ \hline
 2&
 -336 & 2192 & -2192 & -3801 & 6315 & -2178 \\ \hline
 3&
 -221 & 2286 & -2286 & -7666 & 11385 & -3498 \\ \hline
 4&
 -78 & 1456 & -1456 & -8997 & 12364 & -3289 \\ \hline
 5&
 -14 & 575 & -575 & -6642 & 8567 & -1911 \\ \hline
 6&
 -1 & 137 & -137 & -3180 & 3876 & -695 \\ \hline
 7&
 0 & 18 & -18 & -986 & 1140 & -154 \\ \hline
 8&
 0 & 1 & -1 & -191 & 210 & -19 \\ \hline
 9&
 0 & 0 & 0 & -21 & 22 & -1 \\ \hline
 10&
 0 & 0 & 0 & -1 & 1 & 0 \\ \hline
\end{array}
\right];
\end{equation*}}

$\hat{\textbf{N}}_{[1,1]}^{{\bf 5_{1}},c=2}:$
\begin{equation*}
\left[
\begin{array}{c|c|c|c|c|c|c}
g \backslash Q &10&12&14&16&18&20\\

 \hline
0 &
 -45 & 0 & 225 & -225 & 0 & 45 \\ \hline
 1&
 -330 & 0 & 1875 & -1875 & 0 & 330 \\ \hline
 2&
 -924 & 0 & 6315 & -6315 & 0 & 924 \\ \hline
 3&
 -1287 & 0 & 11385 & -11385 & 0 & 1287 \\ \hline
 4&
 -1001 & 0 & 12364 & -12364 & 0 & 1001 \\ \hline
 5&
 -455 & 0 & 8567 & -8567 & 0 & 455 \\ \hline
 6&
 -120 & 0 & 3876 & -3876 & 0 & 120 \\ \hline
 7&
 -17 & 0 & 1140 & -1140 & 0 & 17 \\ \hline
 8&
 -1 & 0 & 210 & -210 & 0 & 1 \\ \hline
 9&
 0 & 0 & 22 & -22 & 0 & 0 \\ \hline
 10&
 0 & 0 & 1 & -1 & 0 & 0 \\ \hline
\end{array}
\right];\end{equation*}

$ \bullet$ For $\left[{\bf 3_1}\right]_{5}: \bf{7_1}$\\

{ $\hat{\textbf{N}}_{[2]}^{{\bf 7_{1}},c=1}:$
 \tiny \begin{equation*}
\left[
\begin{array}{c|c|c|c|c|c|c|c}
g \backslash Q &11&13&15&17&19&21&23\\

 \hline
0 &
 -448 & 1449 & -1449 & 448 & -420 & 735 & -315 \\ \hline
 1&
 -3696 & 13356 & -13356 & 3696 & -8330 & 14210 & -5880 \\ \hline
 2&
 -13104 & 55384 & -55384 & 13104 & -68334 & 112847 & -44513 \\ \hline
 3&
 -26300 & 135762 & -135762 & 26300 & -309817 & 492415 & -182598 \\ \hline
 4&
 -33188 & 218242 & -218242 & 33188 & -883168 & 1347710 & -464542 \\ \hline
 5&
 -27692 & 242234 & -242234 & 27692 & -1702584 & 2493764 & -791180 \\ \hline
 6&
 -15640 & 190892 & -190892 & 15640 & -2322846 & 3267944 & -945098 \\ \hline
 7&
 -6003 & 108262 & -108262 & 6003 & -2309450 & 3124379 & -814929 \\ \hline
 8&
 -1541 & 44275 & -44275 & 1541 & -1704376 & 2220055 & -515679 \\ \hline
 9&
 -253 & 12926 & -12926 & 253 & -942952 & 1184039 & -241087 \\ \hline
 10&
 -24 & 2625 & -2625 & 24 & -391967 & 475020 & -83053 \\ \hline
 11&
 -1 & 352 & -352 & 1 & -121705 & 142506 & -20801 \\ \hline
 12&
 0 & 28 & -28 & 0 & -27784 & 31465 & -3681 \\ \hline
 13&
 0 & 1 & -1 & 0 & -4524 & 4960 & -436 \\ \hline
 14&
 0 & 0 & 0 & 0 & -497 & 528 & -31 \\ \hline
 15&
 0 & 0 & 0 & 0 & -33 & 34 & -1 \\ \hline
 16&
 0 & 0 & 0 & 0 & -1 & 1 & 0 \\ \hline
\end{array}
\right];\end{equation*}}

{\tiny  \begin{eqnarray*}
\hat{\textbf{N}}_{[2]}^{{\bf 7_{1}},c=2}&:&\\
&&\left[
\begin{array}{c|c|c|c|c|c|c|c|c}
g \backslash Q &14&16&18&20&22&24&26&28\\

 \hline
 0&
 -105 & 0 & 0 & 735 & -735 & 0 & 0 & 105 \\ \hline
 1&
 -1820 & 0 & 0 & 14210 & -14210 & 0 & 0 & 1820 \\ \hline
 2&
 -12376 & 0 & 0 & 112847 & -112847 & 0 & 0 & 12376 \\ \hline
 3&
 -43758 & 0 & 0 & 492415 & -492415 & 0 & 0 & 43758 \\ \hline
 4&
 -92378 & 0 & 0 & 1347710 & -1347710 & 0 & 0 & 92378 \\ \hline
 5&
 -125970 & 0 & 0 & 2493764 & -2493764 & 0 & 0 & 125970 \\ \hline
 6&
 -116280 & 0 & 0 & 3267944 & -3267944 & 0 & 0 & 116280 \\ \hline
 7&
 -74613 & 0 & 0 & 3124379 & -3124379 & 0 & 0 & 74613 \\ \hline
 8&
 -33649 & 0 & 0 & 2220055 & -2220055 & 0 & 0 & 33649 \\ \hline
 9&
 -10626 & 0 & 0 & 1184039 & -1184039 & 0 & 0 & 10626 \\ \hline
 10&
 -2300 & 0 & 0 & 475020 & -475020 & 0 & 0 & 2300 \\ \hline
 11&
 -325 & 0 & 0 & 142506 & -142506 & 0 & 0 & 325 \\ \hline
 12&
 -27 & 0 & 0 & 31465 & -31465 & 0 & 0 & 27 \\ \hline
 13&
 -1 & 0 & 0 & 4960 & -4960 & 0 & 0 & 1 \\ \hline
 14&
 0 & 0 & 0 & 528 & -528 & 0 & 0 & 0 \\ \hline
 15&
 0 & 0 & 0 & 34 & -34 & 0 & 0 & 0 \\ \hline
 16&
 0 & 0 & 0 & 1 & -1 & 0 & 0 & 0 \\ \hline

\end{array}
\right];
\end{eqnarray*}}

{\tiny \begin{eqnarray*}
\hat{\textbf{N}}_{[1,1]}^{{\bf 7_{1}},c=1}&:&\\
&& \left[
\begin{array}{c|c|c|c|c|c|c|c}
g \backslash Q &11&13&15&17&19&21&23\\
 \hline
0 &
 -336 & 1099 & -1099 & 336 & -364 & 637 & -273 \\ \hline
 1&
 -2380 & 8841 & -8841 & 2380 & -6370 & 10829 & -4459 \\ \hline
 2&
 -7182 & 31934 & -31934 & 7182 & -46228 & 75803 & -29575 \\ \hline
 3&
 -12144 & 67924 & -67924 & 12144 & -185601 & 291928 & -106327 \\ \hline
 4&
 -12739 & 94119 & -94119 & 12739 & -467545 & 704067 & -236522 \\ \hline
 5&
 -8673 & 89146 & -89146 & 8673 & -793000 & 1143506 & -350506 \\ \hline
 6&
 -3892 & 59109 & -59109 & 3892 & -945778 & 1307368 & -361590 \\ \hline
 7&
 -1141 & 27664 & -27664 & 1141 & -815082 & 1081557 & -266475 \\ \hline
 8&
 -210 & 9086 & -9086 & 210 & -515698 & 657799 & -142101 \\ \hline
 9&
 -22 & 2047 & -2047 & 22 & -241088 & 296010 & -54922 \\ \hline
 10&
 -1 & 301 & -301 & 1 & -83053 & 98280 & -15227 \\ \hline
 11&
 0 & 26 & -26 & 0 & -20801 & 23751 & -2950 \\ \hline
 12&
 0 & 1 & -1 & 0 & -3681 & 4060 & -379 \\ \hline
 13&
 0 & 0 & 0 & 0 & -436 & 465 & -29 \\ \hline
 14&
 0 & 0 & 0 & 0 & -31 & 32 & -1 \\ \hline
 15&
 0 & 0 & 0 & 0 & -1 & 1 & 0 \\ \hline
\end{array}
\right];
\end{eqnarray*}}

{\tiny \begin{eqnarray*}
~~~~~~~~~~~~~~~~~~&&\hat{\textbf{N}}_{[1,1]}^{{\bf 7_{1}},c=2}:\\
~~~~~~~~~~~~~~~~~&&\left[
\begin{array}{c|c|c|c|c|c|c|c|c}
g \backslash Q &14&16&18&20&22&24&26&28\\
 \hline
 0&
 -91 & 0 & 0 & 637 & -637 & 0 & 0 & 91 \\ \hline
1 &
 -1365 & 0 & 0 & 10829 & -10829 & 0 & 0 & 1365 \\ \hline
 2&
 -8008 & 0 & 0 & 75803 & -75803 & 0 & 0 & 8008 \\ \hline
 3&
 -24310 & 0 & 0 & 291928 & -291928 & 0 & 0 & 24310 \\ \hline
 4&
 -43758 & 0 & 0 & 704067 & -704067 & 0 & 0 & 43758 \\ \hline
 5&
 -50388 & 0 & 0 & 1143506 & -1143506 & 0 & 0 & 50388 \\ \hline
 6&
 -38760 & 0 & 0 & 1307368 & -1307368 & 0 & 0 & 38760 \\ \hline
 7&
 -20349 & 0 & 0 & 1081557 & -1081557 & 0 & 0 & 20349 \\ \hline
 8&
 -7315 & 0 & 0 & 657799 & -657799 & 0 & 0 & 7315 \\ \hline
 9&
 -1771 & 0 & 0 & 296010 & -296010 & 0 & 0 & 1771 \\ \hline
 10&
 -276 & 0 & 0 & 98280 & -98280 & 0 & 0 & 276 \\ \hline
 11&
 -25 & 0 & 0 & 23751 & -23751 & 0 & 0 & 25 \\ \hline
 12&
 -1 & 0 & 0 & 4060 & -4060 & 0 & 0 & 1 \\ \hline
 13&
 0 & 0 & 0 & 465 & -465 & 0 & 0 & 0 \\ \hline
 14&
 0 & 0 & 0 & 32 & -32 & 0 & 0 & 0 \\ \hline
 15&
 0 & 0 & 0 & 1 & -1 & 0 & 0 & 0 \\ \hline
\end{array}
\right].
\end{eqnarray*}}
\end{document}